\documentclass[final,5p,times,twocolumn]{elsarticle}

\usepackage{lineno}
\usepackage{graphicx}
\usepackage{amssymb}
\usepackage{pdflscape}
\usepackage[english]{babel}
\usepackage[dvipsnames,svgnames,x11names]{xcolor}
\usepackage{booktabs,tabularx}
\usepackage{multirow}
\usepackage{subfig} 
\usepackage[hyphens]{url}
\usepackage{hyperref}
\usepackage{amsmath}
\usepackage{commath}
\usepackage[english]{babel}
\usepackage[ruled]{algorithm2e}
\SetEndCharOfAlgoLine{}
\usepackage[leftmargin=6em,rightmargin=12em,indentfirst=false]{quoting}
\usepackage{siunitx}
\usepackage{comment} 

\usepackage[final]{changes}
\usepackage{float}

\usepackage{lineno}
\usepackage{tikz}
\usepackage{comment} 
\usepackage[export]{adjustbox}

\usepackage{graphicx}
\usepackage[utf8]{inputenc}
\usepackage[export]{adjustbox}
\usepackage{wrapfig}
\usepackage{dcolumn}

\makeatletter
\newcolumntype{T}[3]{>{\textfont0=\the@{#1}{#2}{#3}}c<{\DC@end}}
\makeatother

\usepackage{pgfplots}
\pgfplotsset{width=10cm,compat=1.9}

\usepackage{array}

\newcolumntype{L}[1]{>{\raggedright\let\newline\\\arraybackslash\hspace{0pt}}m{#1}}
\newcolumntype{C}[1]{>{\centering\let\newline\\\arraybackslash\hspace{0pt}}m{#1}}
\newcolumntype{R}[1]{>{\raggedleft\let\newline\\\arraybackslash\hspace{0pt}}m{#1}}

\usepackage{subfiles}

\usepackage{todonotes}

\journal{Science and Technology for the Built Environment}
\begin{document}
	
	\begin{frontmatter}
		\title{The ASHRAE Great Energy Predictor III competition: Overview and results}
		
        \author{Clayton Miller$^{a,*}$, Pandarasamy Arjunan$^b$, Anjukan Kathirgamanathan$^c$, Chun Fu$^a$, Jonathan Roth$^{a,d}$, June Young Park$^e$, Chris Balbach$^f$, Krishnan Gowri$^g$, Zoltan Nagy$^e$, Anthony Fontanini$^h$, Jeff Haberl$^i$}

		\address{$^a$Building and Urban Data Science (BUDS) Lab, Dept. of Building, School of Design and Environment (SDE), National University of Singapore (NUS), Singapore}
		\address{$^b$Berkeley Education Alliance for Research in Singapore (BEARS), Singapore}
		\address{$^c$UCD Energy Institute, University College Dublin, Ireland}
		\address{$^d$Urban Informatics Lab, Department of Civil and Environmental Engineering, Stanford University, CA, USA}
		\address{$^e$Intelligent Environments Lab (IEL), Department of Civil, Architectural and Environmental Engineering, Cockrell School of Engineering, The University of Texas at Austin, TX, USA}
		\address{$^f$Performance Systems Development of NY, LLC., NY, USA}
		\address{$^g$Intertek Building Science Solutions, WA, USA}
		\address{$^h$National Renewable Energy Laboratory (NREL), CO, USA}
		\address{$^i$Energy Systems Laboratory, Dept. of Architecture, Texas A\&M University, TX, USA}
        \address{$^*$Corresponding Author: clayton@nus.edu.sg, +65 81602452}

%
\begin{abstract}
In late 2019, ASHRAE hosted the Great Energy Predictor III (GEPIII) machine learning competition on the Kaggle platform. This launch marked the third energy prediction competition from ASHRAE and the first since the mid-1990s. In this updated version, the competitors were provided with over 20 million points of training data from 2,380 energy meters collected for 1,448 buildings from 16 sources. This competition's overall objective was to find the most accurate modeling solutions for the prediction of over 41 million private and public test data points. The competition had 4,370 participants, split across 3,614 teams from 94 countries who submitted 39,403 predictions. In addition to the top five winning workflows, the competitors publicly shared 415 reproducible online machine learning workflow examples (notebooks), including over 40 additional, full solutions. This paper gives a high-level overview of the competition preparation and dataset, competitors and their discussions, machine learning workflows and models generated, winners and their submissions, discussion of lessons learned, and competition outputs and next steps. The most popular and accurate machine learning workflows used large ensembles of mostly gradient boosting tree models, such as LightGBM. Similar to the first predictor competition, preprocessing of the data sets emerged as a key differentiator. 
\end{abstract}

\end{frontmatter}

\section{Introduction}
Reducing the overall energy consumption and associated greenhouse gas emissions in the building sector is essential for meeting our future sustainability goals. One promising technology adoption is an energy metering infrastructure, which has been widely deployed around the world. The new paradigm shift of building energy data collection from interval meters, or data loggers, has generated unprecedented amounts of data. Analysis with a data-driven approach has provided informative insights into the built environment \citep{Miller2015-ia, Park2019-ok}. Among various applications, building energy use prediction and forecasting using machine learning models has been a widely explored topic of research since the early 1990s. Research publications in machine learning applied to building performance prediction have steadily been increasing over the last thirty years. One of the first review papers by Zhao and Magoules explained the prediction methods of building energy consumption which included engineering models, statistical models, and machine learning models. One of the conclusions of this early review was that artificial neural networks (ANN) and support vector machines (SVM) could give a highly accurate prediction, without requiring descriptive building information, as long as they have sufficient historical data \citep{Zhao2012-dn}. More recently, Wang and Srinivasan reviewed artificial intelligence-based building energy use prediction studies. They specifically compared multiple linear regression, ANN, SVM, and ensemble methods. Recent work has focused on using ensembles of models, which are large networks of machine learning techniques that reduce errors through diversity. Although ensemble models are computationally tricky, they usually outperformed other algorithms \citep{Wang2017-yn}. Amasyali and El-Gohary outlined several papers describing methods to predict energy consumption in buildings using features such as weather, occupancy, and schedule \citep{Amasyali2018-wj}. In contrast to supervised learning, Bourdeau et al. identified that researchers also utilized unsupervised, reinforcement, and transfer learning approaches to predict future building energy consumption \citep{Bourdeau2019-xt}.

Despite the overwhelming amount of research in data-driven energy prediction, it is difficult to compare different prediction methods against each other \citep{Miller2019-sg}. Each researcher who implements their technique on a small dataset for a single or small group of buildings is essentially creating a custom-built or bespoke modeling process that has been optimized for the characteristics of that particular context. To address this issue, there have been two critical studies in the last decade using 400 buildings \citep{Granderson2015-ms} and 537 buildings \citep{Granderson2016-wq} building upon a previously generated methodology \citep{Granderson2014-xl} to compare the performance of several prominent energy prediction techniques for measurement and verification (M\&V). While making progress in the field, these studies restricted models tested to only a set of techniques chosen by the authors that are already used on building energy data. The Energy Valuation Organization (EVO) now hosts a platform based on these studies that allows users to test their approaches on 367 buildings\footnote{\url{https://mvportal.evo-world.org/}}. Such work has provided objective evidence of the effectiveness of various techniques and has provided the foundation for further work in automating the M\&V process \citep{Granderson2017-lm}. These studies and the effort associated with them are important, but the number of techniques tested is still relatively low and the data used in these studies are not entirely available in an open access way to the broader research community. These aspects restrict the ability to compare the numerous new machine learning models and techniques rapidly being developed by the modern data science movement.

The built environment is not the only domain that has the challenge of generalizability of models, and many of them have created data sets and competitions to respond. For example, the research field of computer vision has been developed extensively by strong open datasets, e.g., CIFAR-10 \citep{Krizhevsky2009-ou}, Cityscapes \citep{Cordts2016-xv}, Fashion-MNIST \citep{Xiao2017-rd}, and ImageNet \citep{Russakovsky2015-iu}. Lessons learned from the computer vision research domain include the fact that generalizability and scalability of solutions can be investigated through large machine learning competitions. 

This paper describes the development and results of the ASHRAE Great Energy Predictor III (GEPIII) competition held in late 2019. Since the first launch in 1993, the ASHRAE Great Energy Predictor competition series has been the foundation for crowdsourced machine learning benchmarking for time series data related to the building and construction industry. This paper will show that this latest competition effort has crowdsourced a significant amount of machine learning modeling knowledge for determining the best-performing methods for the hourly energy prediction of commercial buildings and tutorials for building scientists who would like to learn these techniques.

\subsection{Machine learning competitions as a means of crowdsourcing and benchmarking prediction models}
Machine learning competitions generally consist of providing the same dataset, prediction objectives, constraints, and performance metrics to a pool of contestants, intending to determine who can develop the best workflow/model/algorithm to achieve the most accurate prediction results. Hundreds of machine learning competitions have been held since the early 2000s when the internet made it possible to crowdsource the machine learning process and provide instantaneous results to participants. Several platforms exist, with the most popular hosting several competitions with substantial monetary incentives such as the Zillow Home Value Prediction competition that had total prizes of \$1.2 million USD\footnote{\url{https://www.kaggle.com/c/zillow-prize-1}}. Over 25 years have passed since the Great Energy Predictor Competition II (also known as the Predictor Shootout II); however, the models and lessons learned from that competition are still being used in the research community. A significant motivation for this competition was to adopt the latest developments in machine learning since the last competition into the community of building researchers.

The Great Energy Predictor Shootout I contest, held in 1993, was hosted by Jeff Haberl and Jan Kreider, with support from ASHRAE TC 4.7 and TC 1.5 \citep{Kreider1994-dn}. For this contest, more than 150 participants were provided a four-month-long training dataset containing hourly records of chilled water, hot water, and whole building electricity usage \emph{for a single institutional building}. Detailed weather data, including solar radiation measurements, were also provided. Contestants were asked to construct different models to predict the energy use for the two months that followed the data set and a time-independent model predicting solar flux. Submissions were sent to the contest organizers via shipping and mailing services. These data were compared to the actual measured energy usage and solar measurements and scored using the Coefficient of Variation (CV) metric. While monetary prizes were not awarded, six winners were recognized in a variety of publications and presentations. The top winners used Bayesian nonlinear modeling \citep{MacKay1996-ht}, artificial neural networks \citep{Ohlsson1994-vl, Stevenson1994-ps}, generalized nonlinear regression with an ensemble of neural networks \citep{Feuston1994-mm}, and piecewise linear regression \citep{Iijima1994-gj}. After the competition, the specifics of the dataset were revealed to the contestants, and papers were written to detail their modeling efforts. In addition, ASHRAE published a special issue that contained the papers and data sets from the competition \citep{Haberl1999-tq}.

A second Great Energy Predictor Shootout II contest was conducted in 1994 and hosted by Jeff Haberl and Sabaratnan Thamilseran \citep{Haberl1998-du}. This six-month competition leveraged an anonymous FTP (file transfer protocol) site, where 50 downloads of the instructions and data were recorded. Contestants were provided with detailed sub-metered datasets collected from \emph{two buildings} that had recently received energy savings retrofits. For each building, the competition utilized sets of measured hourly pre- and post-retrofit data. Contestants were asked to construct models to predict data that had been carefully and purposefully removed from both the pre- and post-retrofit periods. A single set of predictions were submitted to the organizers, who compared them to the hold-out data and calculated the accuracy metrics of the hourly coefficient of variation (CV-RMSE) and mean bias error (MBE). The Predictor Shootout II competition had 47 submissions, with only four submissions meeting all the submission requirements. While monetary prizes were not awarded, the four compliant submissions were recognized in a variety of publications and presentations \citep{Katipamula1996-et, Chonan1996-rz, Jang1997-zm, Dodier2004-pg}. 

\section{Pre-Competition process}
The planning of the third competition kicked off in January 2018 at the ASHRAE Winter Meeting in Chicago, IL. Following this, the ASHRAE Technical Committee (TC) 4.7 discussed creating a machine learning competition. The logistics and technical leaders of the competition began a discussion in February 2018, where contest scope, goals, timeline, and resources needed were planned. The plan was initially presented to the ASHRAE Technical Activities Committee (TAC) in June 2018 at the ASHRAE summer meeting in Houston, TX. While the TAC indicated general support and encouragement, commitment for the required monetary resources was not ensured. Nevertheless, the organizing committee continued expanding efforts - recruiting potential \emph{data donors} and refining expectations. The organizing committee identified several possible hosting platforms for the data competition. The team selected Kaggle\footnote{\url{https://www.kaggle.com/}}, having over ten years of competition hosting experience and more than one million registered users, as the \emph{preferred} host for the data competition. Initial contacts with Kaggle were made in September of 2018. A more detailed proposal was presented to the TAC in January 2019 at the ASHRAE Atlanta Winter meeting. The TAC, along with ASHRAE Tech Council, recommended that the detailed proposal be presented to the ASHRAE Research Activities Committee (RAC) - to be funded as an \emph{Unsolicited Research Proposal}. At the June 2019 ASHRAE Summer meeting, the RAC voted to approve the data competition and fund the monetary awards, allowing the data competition to move forward. Soon after, a complete competition dataset was shared with Kaggle, who maintained strict data vetting and acceptance processes. Throughout July and August, the organization committee worked closely with Kaggle to finalize the dataset and present competition rules to ASHRAE to approve. ASHRAE formalized their support by signing a contract with Kaggle in September of 2019. Kaggle formally launched the ASHRAE sponsored data competition on October 5, 2019, with a scheduled ending of December 19, 2019. 

\subsection{Dataset development}
The creation of the competition dataset started in March 2018 through initial discussions with the various stakeholders by the technical lead. This effort continued through 2018 up to May of 2019. This data collection process's primary goal was to create the largest and most diverse dataset possible to challenge the contestants to create the most generalizable models for the benefit of the energy prediction research community. Requests for data donors were made at various ASHRAE conferences and through the technical team members. The datasets were collected from publicly available sources that are freely available online and from closed systems that required extraction by the facility management teams from many of the data donor locations. 

\begin{figure*}[ht!]
\centering
\includegraphics[width=\textwidth,height=\textheight,keepaspectratio]{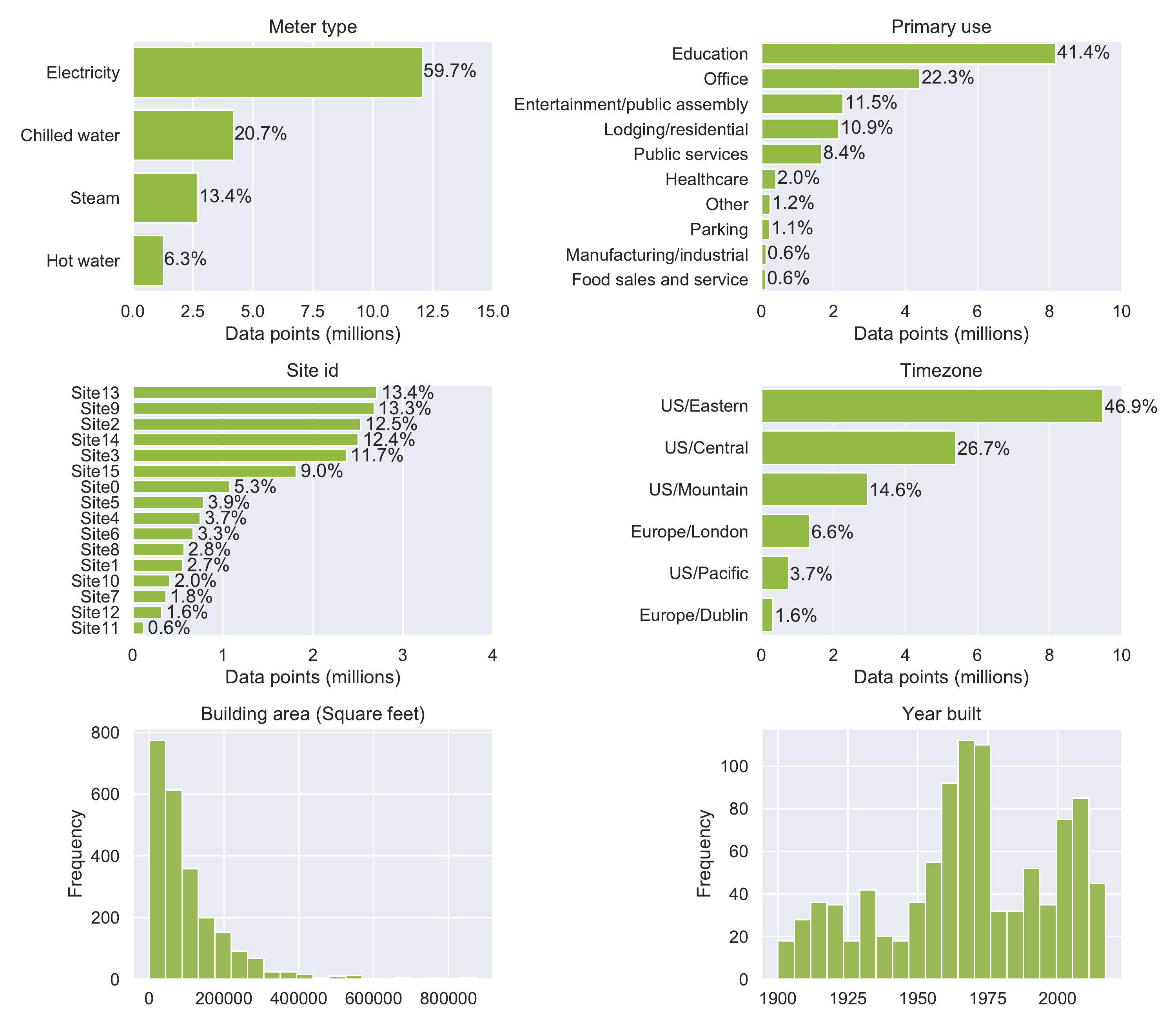}
\caption{Overview of the data set curated for the GEPIII competition (clockwise from upper left): type of energy being measured (upper left); the primary use type of the buildings in the data set (upper right); the time zone in which the buildings are located (middle right); the year the buildings were constructed (lower right); the gross floor area in sq. ft. of the buildings (lower left); and a breakdown of the amount of data collected from the 16 sites (middle left).}
\label{fig1-data}
\end{figure*}

Throughout the data collection process, the dataset grew to a total of 61,910,200 energy measurements taken from 16 sites worldwide. These data are hourly measurements taken from energy metering systems at locations that had data from January 1, 2016, to December 31, 2018. Figure \ref{fig1-data} illustrates a breakdown of crucial metadata features of the data set.  A majority of the data sites were universities; therefore, the most common building type is education. The primary use type of buildings corresponds to the categories found in the EnergyStar building benchmarking system. Around 73\% of the data were from buildings on university campus sites (1058 buildings) and the remaining 27\% of the sites were from city-wide municipal and healthcare building repositories (390 buildings). Some of the universities from which data were collected include the University of Central Florida, University College London, Arizona State University, University of California at Berkeley, University of Texas at Austin, Carleton University, University College Dublin, Princeton University and Cornell University. The municipal building sites included Washington DC, Cardiff, UK, and Ottawa, Canada. These sites as well as others that remain anonymous are found in a complimentary publication \citep{Miller2020-hc}.

Coincident weather data was provided to the contestants for each of these sites. Minimal data cleaning and processing were conducted on the data for the competition as the technical committee wanted conditions for the competitors to be as close to a real-world scenario in which data cleaning and preprocessing is an integral component of the winning contestants’ solutions. Details of the cleaning and preparation of the data set, the sources from which the data was collected, and other information on data use can also be found in the publication that outlines the open release of much of the competition data set \citep{Miller2020-hc}. 

\subsection{Competition objective}
\begin{figure*}[ht!]
\centering
\includegraphics[width=\textwidth,height=\textheight,keepaspectratio]{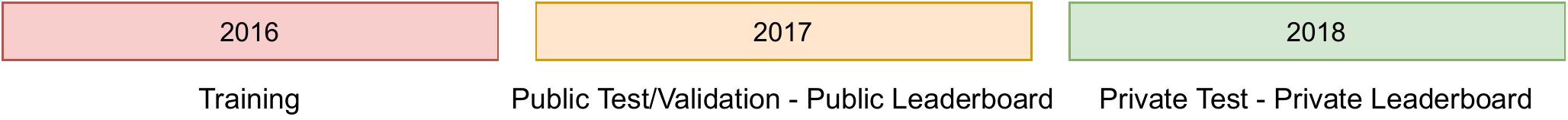}
\caption{Split between training, public test/validation and private test data}
\label{fig2-split}
\end{figure*}

The context presented to the contestants focused on using regression-based machine learning to predict the energy savings of a retrofit in the measurement and verification (M\&V) process. Assessing the value of energy efficiency improvements can be challenging as there is no economically viable way to truly know how much energy a building would have used without the upgrades. Therefore, the best solution is to build counterfactual models that predict the building’s pre-retrofit energy use rate in the post-retrofit period. Once a building is overhauled, the new (lower) energy consumption is compared against modeled values for the original building to calculate the retrofit's savings. More accurate models could support better market incentives and enable lower-cost financing. This competition challenged the contestants to build these counterfactual models across four energy types based on historical energy usage rates and observed coincident weather. Prediction of hourly data in this context is not always standard; however, this aspect was seen as a means of increasing the challenge for the participants. To create the machine learning challenge, each of the three years of data was divided into its own data set, as seen in Figure \ref{fig2-split}. 

\subsubsection{Training data set}
The first year (2016) of hourly energy meter data was provided to the machine learning contestants as the \emph{training} data to be used for model training alongside various hourly weather data streams. The contestants were tasked with preprocessing these data, creating multiple input features, and training machine learning modeling frameworks to predict the second and third years of data. 

\subsubsection{Public test/validation data set - the public leaderboard}
The second-year (2017) of meter data was not provided to the contestants. It was assigned as the \emph{public test/validation} data set for an incremental indication of progress in the machine learning process during the competition through the \emph{public leaderboard}. The contestants were not notified of where the exact divisions were between the \emph{public test/validation} and \emph{private test} data sets, but they were aware that there would be this division and that there were separate scoring and leaderboards for each. This data set included data from all the buildings and meters, \emph{including those from public data sources}. The contestants were not notified that a portion of the data sets was public, but the competition planning team understood that some of these data would be discovered and as a result were shared openly within the contestant pool in the discussion forum. Public data sources were considered \emph{external data} and were allowed to be used in the development of the solution, as long as these sources were disclosed to the other contestants. We call this data set \emph{public test/validation} as it was assumed that the public data could be used as a \emph{validation} set to help refine strategies. 

\subsubsection{Private test data set - the private leaderboard}
The third-year (2018) was also withheld from the contestants and was assigned to the final \emph{private test} data set that was used to calculate the score that was used to determine the winning individuals and teams. This final score was withheld from the contestants during the competition and was released as the \emph{private leaderboard} after the competition was concluded. The calculation of this score \emph{did not include} data from the public data sources; therefore, the contestants' discovery and use of these data had no impact on this score beyond providing more training and validation data. All public data was eventually shared on the discussion forum and all contestants had access. An overview of which sites were publicly available can be found in the associated open data publication mentioned earlier \citep{Miller2020-hc}. During the competition, one of the sites that were previously thought to be non-public (by the operations team of the data donor) was found to have an API that the contestants found. Therefore, this data set was removed from the \emph{private test data set} to prevent a leakage impact on the \emph{final private test} (private leaderboard) score.

\section{Competition procedures and rules}
\begin{figure*}[ht]
\centering
\includegraphics[width=\textwidth,height=\textheight,keepaspectratio]{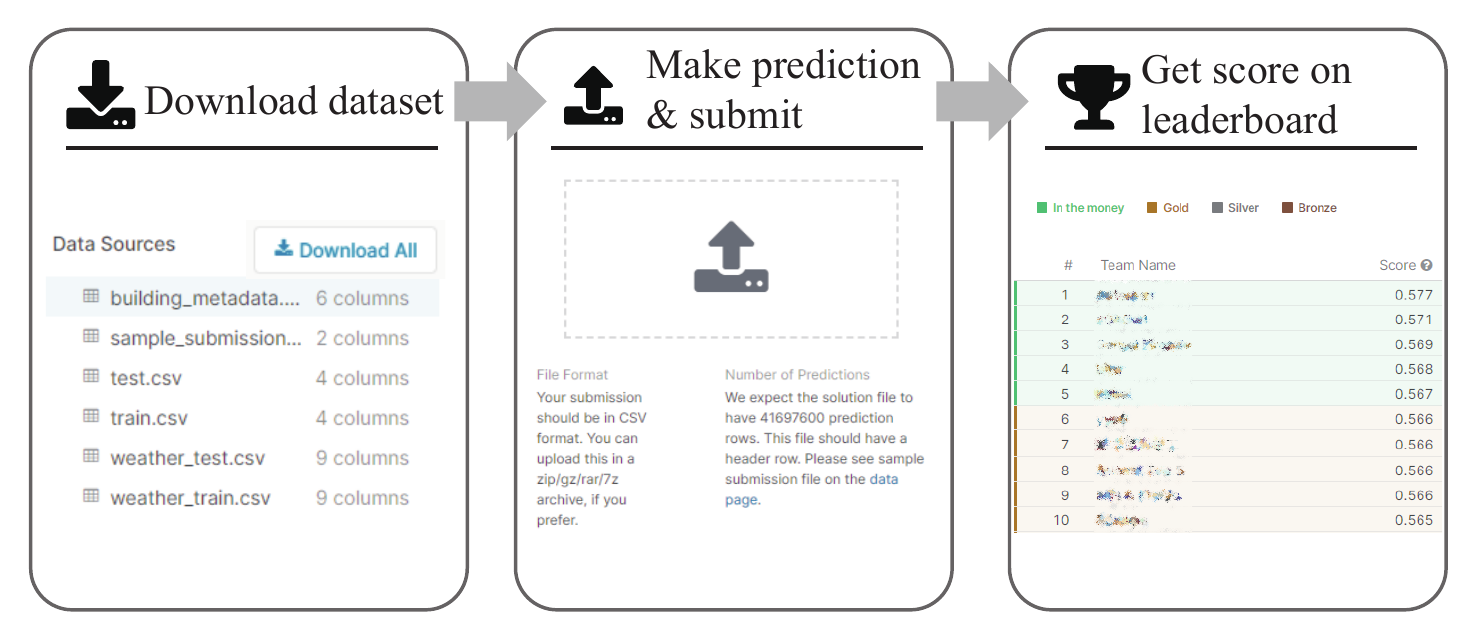}
\caption{Workflow for the competitors to make a single submission.}
\label{fig3-process}
\end{figure*}

To participate, contestants were first asked to create an account or login to the competition platform, find the hosted page, and select that they would like to join the competition. They were then asked whether they accepted the conditions of the competition and upon agreement were provided access to the \emph{training data set}, a metadata file about the buildings and meters, and the associated weather data for the whole three year period. Figure \ref{fig3-process} outlines the process that each participant went through to collect the training data, make and submit their predictions, and then view an automatically created score of their prediction on the \emph{public leaderboard}. 

\subsection{Rules overview}
The planning and technical committee of the competition worked closely with the competition host platform to develop rules and guidelines that would maximize the benefit of the results of the competition by providing a level playing field for the competitors. The following list is a highlight of the relevant rules of the competition:
\begin{itemize}
  \item Participants were not allowed to compete from multiple accounts. 
  \item Participants were allowed to compete individually, or multiple individuals were allowed to collaborate as a team of up to five people. 
  \item The individuals and teams were allowed a maximum of 2 entries per day and were allowed to select two final submissions for judging at the end of the competition.
  \item Participants were allowed to use data other than the competition data to develop and test their models and submissions. However, they were required to ensure the external data is available for use by all participants of the competition for purposes of the competition at no cost to the other participants and post such access to the external data for the participants to the official competition forum before the entry deadline.
  \item The use of automated machine learning tools in the creation of submissions was permitted, but teams that used them were not eligible to win prizes.
  \item Privately sharing code or data outside of teams was not permitted. However, it was okay to share code if made available to all participants on the forums.
\end{itemize}

\subsection{Evaluation metric}

Whenever a contestant submitted a prediction, it was scored using the Root Mean Squared Logarithmic Error (RMSLE). This metric was selected because it is a common and straightforward adaptation of the Root Mean Square Error (RMSE) that reduces the risk that meters with much larger consumption values would unfairly influence the score significantly more than lower consuming meters. Since this competition was hosted as a \emph{non-profit} competition, it removed the technical team's ability to create a custom scoring metric.

The RMSLE is calculated with Equation \ref{eq:metric}:
\begin{equation}
\label{eq:metric}
\epsilon=\sqrt{\frac{1}{n} \sum_{i=1}^{n}\left(\log \left(p_{i}+1\right)-\log \left(a_{i}+1\right)\right)^{2}}
\end{equation}

Where:
\begin{itemize}
    \item $\epsilon$ is the RMSLE value (score),
    \item $n$ is the total number of observations in the (public/private) dataset,
    \item $p_i$ is the prediction of the target,
    \item $a_i$ is the actual target for $i$, and
    \item $log(x)$ is the natural logarithm of $x$.
\end{itemize}

\subsection{Prizes and winners’ obligations}
Upon launch of the competition, the contestants were made aware of the financial rewards of winning the competition. The monetary prizes were supported by ASHRAE and included the following breakdown in US dollars:

\begin{itemize}
    \item 1st place - \$10,000
    \item 2nd place - \$7,000
    \item 3rd place - \$5,000
    \item 4th place - \$2,000
    \item 5th place - \$1,000
\end{itemize}

To claim the prize money, the participants were asked to accept the following obligations when they registered for the competition:
\begin{itemize}
    \item The winners were obligated to deliver to ASHRAE the final model's software code as used to generate the winning submission and associated documentation. The delivered software code was required to fulfill certain documentation guidelines and be capable of generating the winning submission. It must also contain a description of the resources required to build and run the executable code successfully.
    \item Winners were required to grant ASHRAE an Open-Source license to the winning submission and represent that they had the unrestricted right to grant that license.
\end{itemize}

In addition to the prize money, winners were encouraged to attend ASHRAE conferences and participate in workshops, events, etc. In addition to the winner obligations, the winning competitors were asked to create a short video (3-7 minutes) summarizing their solution.

\subsection{Data files}
When the competitors successfully enrolled in the competition they were given access to several raw data files that contained the training dataset as well as several types of metadata that describe each building, and coincident weather data from the individual sites. Table \ref{tab:table-datafiles} outlines the files and associated data columns that were provided to the contestants:

\begin{table*}[ht]
\caption{Overview of the files provided to the contestants.} 

\label{tab:table-datafiles}
\centering

\begin{tabular}{|p{7cm}|p{10cm}|}
    \hline
    \textbf{File Name and Description} & \textbf{File Variables and Short Descriptions}\\
    \hline
    \multirow{4}{7cm}{\texttt{train.csv} - This file includes one year of hourly time-series (8760 samples per meter) for each meter. The date range for this training set is Jan 1 - Dec 31, 2016. These data are used to train the prediction models.} 
    &\texttt{building\_id} - Foreign key for the building metadata.\\
    &\texttt{meter} - The meter id code. Read as 0: electricity, 1: chilledwater, 2: steam, 3: hotwater. Not every building has all meter types.\\
    &\texttt{timestamp} - When the measurement was taken.\\
    &\texttt{meter\_reading} - The target variable. Energy consumption in kWh (or equivalent). Note that this is real data with measurement error, which we expect will impose a baseline level of modeling error.\\
    \hline 
    \multirow{1}{7cm}{\texttt{building\_meta.csv} - This file includes the characteristic data from each building in the competition}
    &\texttt{site\_id} - Foreign key to match with weather.csv.\\
    &\texttt{building\_id} - Foreign key for training.csv.\\
    &\texttt{primary\_use} - Indicator of the primary category of activities for the building based on EnergyStar property type definitions.\\
    &\texttt{square\_feet} - Gross floor area of the building.\\
    &\texttt{year\_built} - Year building was opened.\\
    &\texttt{floor\_count} - Number of floors of the building.\\
    \hline
    \multirow{1}{7cm}{\texttt{weather\_[train/test].csv} - Weather data from a meteorological station as close as possible to the site. The data is for all sites and spans from Jan. 1, 2016 to Dec. 31, 2018.}
    &\texttt{site\_id} - Foreign key to match with the meta.csv file.\\
    &\texttt{air\_temperature} - Degrees Celsius\\
    &\texttt{cloud\_coverage} - Portion of the sky covered in clouds, in oktas\\
    &\texttt{dew\_temperature} - Degrees Celsius\\
    &\texttt{precip\_depth\_1\_hr} - Precipitation in millimeters\\
    &\texttt{sea\_level\_pressure} - Millibar/hectopascals\\
    &\texttt{wind\_direction} - Compass direction (0-360)\\
    &\texttt{wind\_speed} - Meters per second\\
    \hline
    \multirow{1}{7cm}{\texttt{test.csv} - The submission files use row numbers for ID codes in order to save space on the file uploads. test.csv has no feature data; it exists so you can get your predictions into the correct order. The test data submissions span from Jan. 1, 2017 to Dec. 31, 2018.}
    &\texttt{row\_id} - Row id for the submission file\\
    &\texttt{building\_id} - Building id code\\
    &\texttt{cloud\_coverage} - Portion of the sky covered in clouds, in oktas\\
    &\texttt{dew\_temperature} - Degrees Celsius\\
    &\texttt{meter} - The meter id code\\
    &\texttt{timestamp} - Timestamps for the test data period\\
    \hline
\end{tabular}
\end{table*}

\subsection{Timeline}
The competition had the following important dates (all times are 11:59 PM UTC):

\begin{itemize}
    \item Start date when the contestants were first granted access to register, download the data, submit solutions, and form teams: October 15, 2019
    \item Entry deadline for registering for the competition: December 12, 2019
    \item Merger deadline when contestants were last able to form a team: December 12, 2019
    \item End date and final submission deadline: December 19, 2019
\end{itemize}

\section{Overview of the competitors}

\begin{figure*}[ht]
\centering
\includegraphics[width=\textwidth,height=\textheight,keepaspectratio]{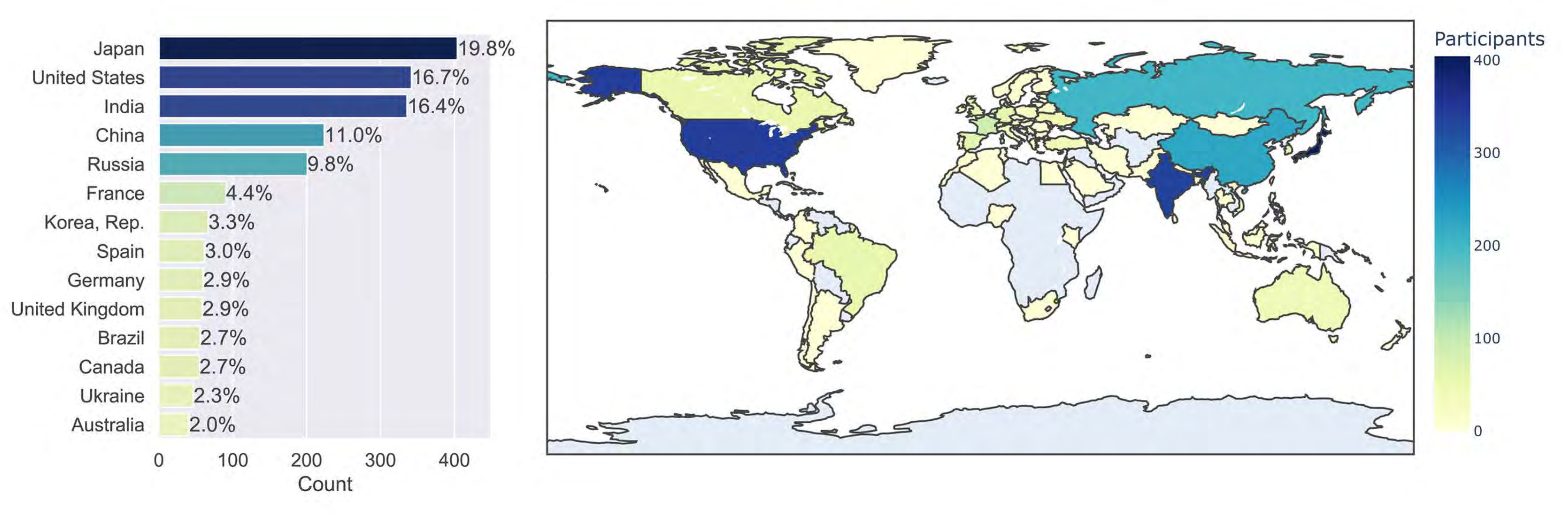}
\caption{Overview of the locations of the 58.1\% of the participants who had a country of origin listed in their profiles. Japan, the USA, India, China, and Russia were the top five countries of origin of the contestants.}
\label{fig4-map}
\end{figure*}

The competition was launched online on October 15, 2019 on the Kaggle website\footnote{\url{https://www.kaggle.com/c/ashrae-energy-prediction}}. During the competition, a total of 4,370 participants took part, comprising a total of 3,614 teams. Of these participants, a total of 2,522 (or 58.1\%) indicated their country of origin. Overall, there were a total of 80 countries represented among these participants. Figure \ref{fig4-map} shows a map of the countries of origin of the 2,522 participants across the 80 countries. The figure visualizes the number of participants from those countries that were represented by more than 1\% of participants, where the \emph{Other Countries} box is the summation of all countries with less than 1\% of participants. 

\begin{figure*}[ht]
\centering
\includegraphics[width=\textwidth,height=\textheight,keepaspectratio]{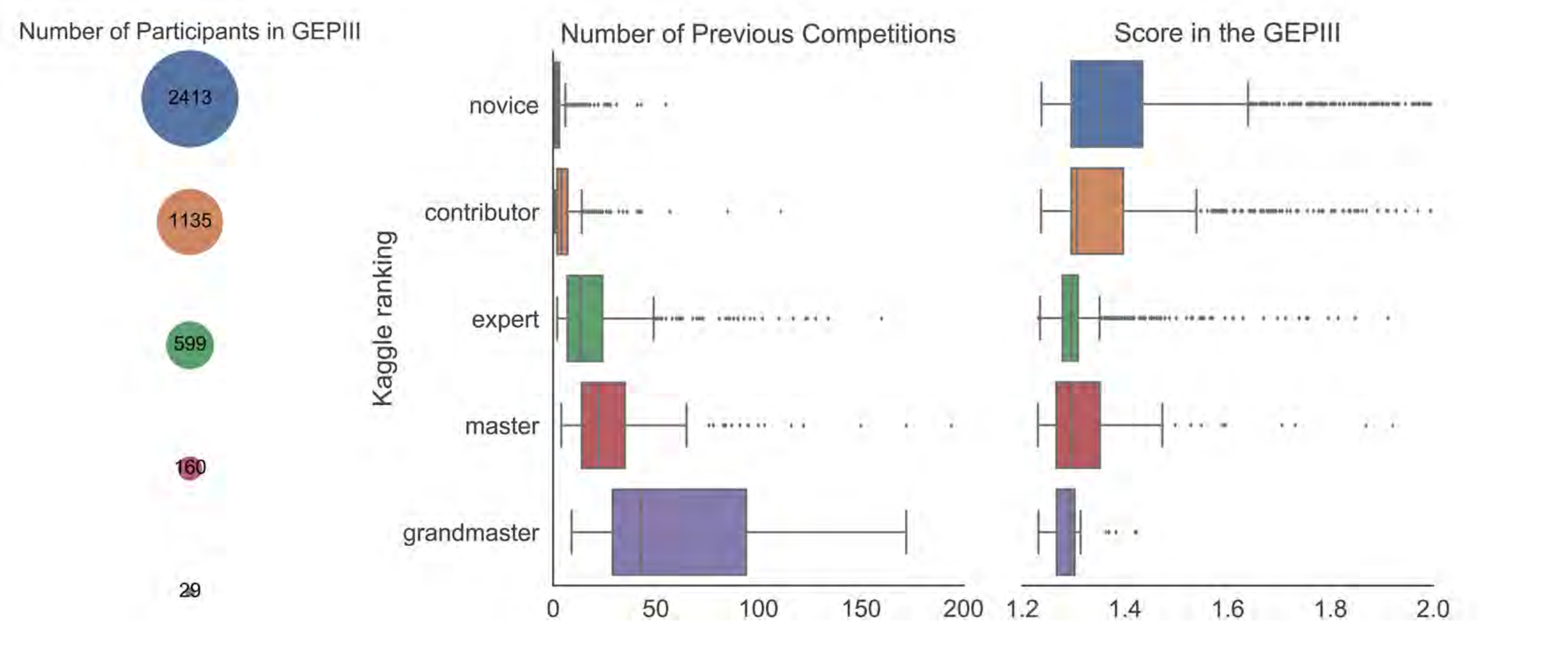}
\caption{Overview of the participants broken down by the total number of participants from each Kaggle expertise category (left), the number of previous competitions they have participated in previously (middle), and ranking based on their final \emph{private test data} scores from the private leaderboard (right).}
\label{fig5-participants}
\end{figure*}

Kaggle has five contestant classification tiers that can be achieved: \emph{Novice}, \emph{Contributor}, \emph{Expert}, \emph{Master}, and \emph{Grandmaster}. Users of the platform start as a \emph{Novice} and can level-up through various thresholds of activity such as competition standings, votes on their analysis and discussion topics, and other activities. Figure \ref{fig5-participants} shows the number of participants in this competition from each Kaggle ranking category, the number of past competitions that they had competed in previously, and a distribution of their private leaderboard scores. Participants with a \emph{Novice} ranking were, by far, the largest group, containing a total of 2,413 people, while there were only a total of 29 \emph{Grandmasters}. Nearly all of the participants with the \emph{Novice} ranking competed in one or fewer past competitions, and they also received, on average, the lowest scores out of all participant types. The trend from this figure shows that high ranking aligns with more past competition experience and lower (i.e., better) scores in this competition. 

\subsection{Discussion board overview}
The Kaggle competition environment also provided a discussion forum for communications between Kaggle and competitors as well as between the competitors themselves. From the start of the competition, participants were encouraged to discuss any aspects of the competition. The metadata collected from this discussion board provides a useful resource to analyze various issues of the competition, such as sentiment, level of engagement, participation, and critical topics that emerged from the competition. The analysis performed extracted this metadata from the discussion board from the start of the competition to the cut-off date of December 28, 2019, a full week after the final submission deadline. The data collected during this time related to 293 discussion posts and 2,508 comments on the posts involving 713 unique competitors. 

\begin{figure*}[ht]
\centering
\includegraphics[width=\textwidth,height=\textheight,keepaspectratio]{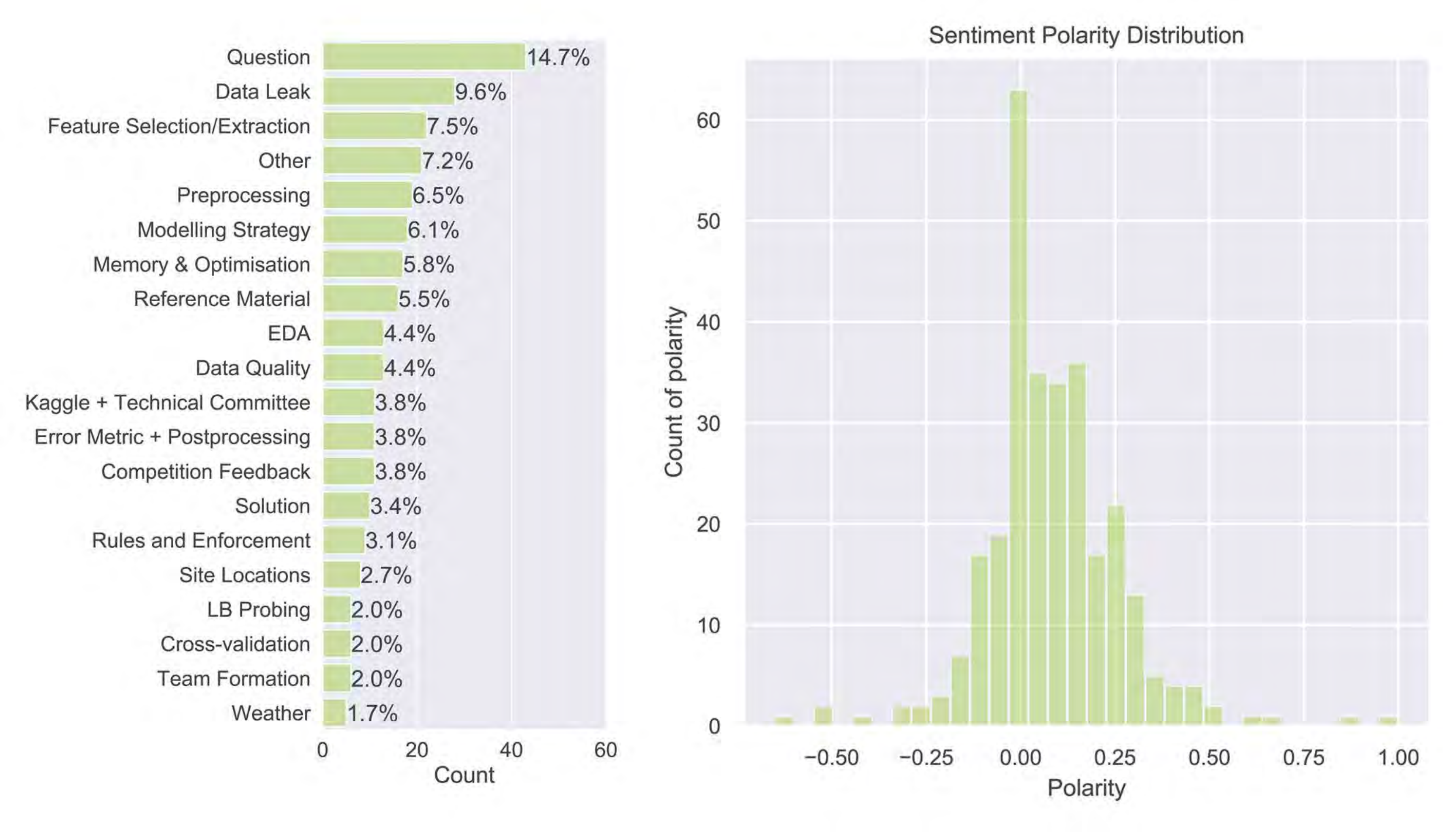}
\caption{Overview of the discussion board topics posted (left) and a histogram of the sentiment polarity analysis of the discussion posts (right). The sentiment of the discussions were slightly more positive than average.}
\label{fig6-discuss} 
\end{figure*}

Each discussion post was manually tagged with a label characterizing the nature of the post, and a visualization of these results is presented in Figure \ref{fig6-discuss} (left). The largest categories of topics focused on questions, discussion of the public data sources, feature extraction and selection, preprocessing, and modeling strategy. Some of the issues exhibited positive tones, predominantly those focused on cooperation between the contestants, while others, especially those focused on the use of the public data for the \emph{public test/validation data} set, were more negative.

To classify the overall tone of the discussion board, all text content from the discussion posts and comments were extracted, and sentiment analysis was carried out using the TextBlob\footnote{TextBlob: Simplified Text Processing. \url{https://textblob.readthedocs.io/en/dev/}} package in Python. The sentiment polarity of every post was extracted, and a histogram of the results is presented in Figure \ref{fig6-discuss} (right). A value of \texttt{+1.0} represents the maximum positive sentiment whereas a value of \texttt{-1.0} represents the most negative sentiment. The figure illustrates that the sentiment polarity distribution is right-shifted (with a mean greater than 0), indicating a higher than average positive sentiment to the competition in general. These results are presented to understand a data-driven perspective of the subjective nature of the competition and are not meant as a comparison to other online competitions.

\section{Overview of the top five and medal-winning teams or individuals}
As previously mentioned, the competition's primary goal was to find the top-performing machine learning workflows as evaluated by the performance metrics on the \emph{private test data set} (private leaderboard). The top teams and individuals were able to create the most accurate solutions and therefore were able to claim the five levels of prize money. In return for the prizes, they were obligated to share the details of their machine learning workflows. These solutions include several steps that are interesting in setting each solution apart from the rest of the leaderboard. Table \ref{table-winners}  gives a high-level overview of those top five solutions with a brief description of the following categories:

\begin{itemize}
    \item Preprocessing Strategy - This step includes the various transformations that clean and prepares the data for the modeling process. The removal of outliers and anomalous behavior that might influence the models are usually included in this process.
    \item Feature Strategy - Features are the aggregations of the time-series data that are used to train the models. These features are extracted from the available training data, including the weather data. A better selection or engineering of useful features enables better-informed and more efficient models.
    \item Modeling Strategy - The selection and tuning of machine learning models is a major component of the ability for the solution to achieve accuracy. The top five winners were consistent in that they used an array of model types known as an ensemble. A more detailed description of the different models is found in the Discussion Section.
    \item Postprocessing Strategy - After the models were created, trained, and the test data predicted, many of the contestants added adjustments to make slight corrections before submission.
\end{itemize}

An open-source repository has been made available on Github to see this documentation and reproduce the solutions\footnote{\url{https://github.com/buds-lab/ashrae-great-energy-predictor-3-solution-analysis}}. This repository has a 3-7 minute video for each of the winning solutions. The technical team had the opportunity to discuss the solutions that were submitted from each of the winning teams or individuals on a one-hour conference call that included the winners, many of the technical team members, and the Kaggle project leader. Some of the anecdotal information from those calls is included in the solution explanations. Several of the winners used publicly available data in their solution diagrams and in many cases these data are labelled as \emph{leaked data}.

\begin{table*}[ht!]
\caption{Overview of the winning teams and individuals according to rank, team composition (Team), Final private test score - Private leaderboard score (Score), Preprocessing strategy overview (Preprocess), Simplified feature explanation (Features), Modelling strategy (Modeling), and Post-processing strategy (Postprocess)}
\label{table-winners}
\centering

\begin{tabular}{|p{0.1cm}|p{3.2cm}|p{0.65cm}|p{2.7cm}|p{2.7cm}|p{2.7cm}|p{2.7cm}|}
\hline
\textbf{\#} & \textbf{Team} & \textbf{Score} & \textbf{Preprocess} & \textbf{Features} & \textbf{Modeling} & \textbf{Postprocess} \\
\hline
1 & Matthew Motoki and Isamu Yamashita  & 1.231 & Removed anomalies in meter data and imputed missing values in weather data & 28 features engineered including holidays & LightGBM, CatBoost, and MLP models trained on different subsets of the training and public data & Ensembled the model predictions using weighted generalized mean. \\
\hline
2 & Rohan Rao, Anton Isakin, Yangguang Zang, and Oleg Knaub & 1.232 & Visual analytics and manual inspection & 35 features using raw meter data, temporal, building metadata, simple statistical features of weather. & XGBoost, LightGBM,
Catboost, and Feed-forward Neural Network models trained on different subset of the training set & Weighted mean. (different weights were used for different meter types) \\
\hline
3 & Xavier Capdepon & 1.234 & Eliminated 0s in the same period in the same site & 21 features including raw data, weather, and various meta data & Keras CNN, LightGBM and Catboost & Weighted average \\
\hline
4 & Jun Yang & 1.235 & Deleted outliers during the training phase & 23 features including raw data, aggregate, weather lag features, and target encoding. Features are selected using sub-training sets. & XGBoost (2-fold, 5-fold) and Light GBM (3-fold) & Ensembled three models. Weights determined using publicly available data from the \emph{public test/validation} data set. \\
\hline
5 & Tatsuya Sano, Minoru Tomioka, and Yuta Kobayashi & 1.237 & Dropped long streaks of constant values and zero target values. & 10 features using target encoding using percentile and proportion and used the weather data temporal features & LightGBM in two steps to identify model parameters on a subset and then train on the whole set. & Weighted average.
\\
\hline
\end{tabular}

\end{table*}

\subsection{First-place solution - Group-based ensembles using CatBoost, LightGBM, and MLP}

The first-place winning team of the competition was made up of Matthew Motoki of Honolulu, HI, USA, a Senior Data Scientist at Iterable, and Isamu Yamashita of Yokohama, Japan, a Machine Learning Researcher at Cannon. Their \emph{final private test score} (private leaderboard) was 1.231 and their \emph{public test/validation set} score (public leaderboard) was 0.938 with a rank of 14. Matthew mentioned in the overview call that their solution was influenced by an experience he had as a data science consultant to an engineering project that focused on water meter data prediction. Both Matthew and Isamu are at the \emph{Master} level on the Kaggle platform, signifying that they each have extensive machine learning competition experience.

\begin{figure}[ht]
\centering
\includegraphics[width=\linewidth]{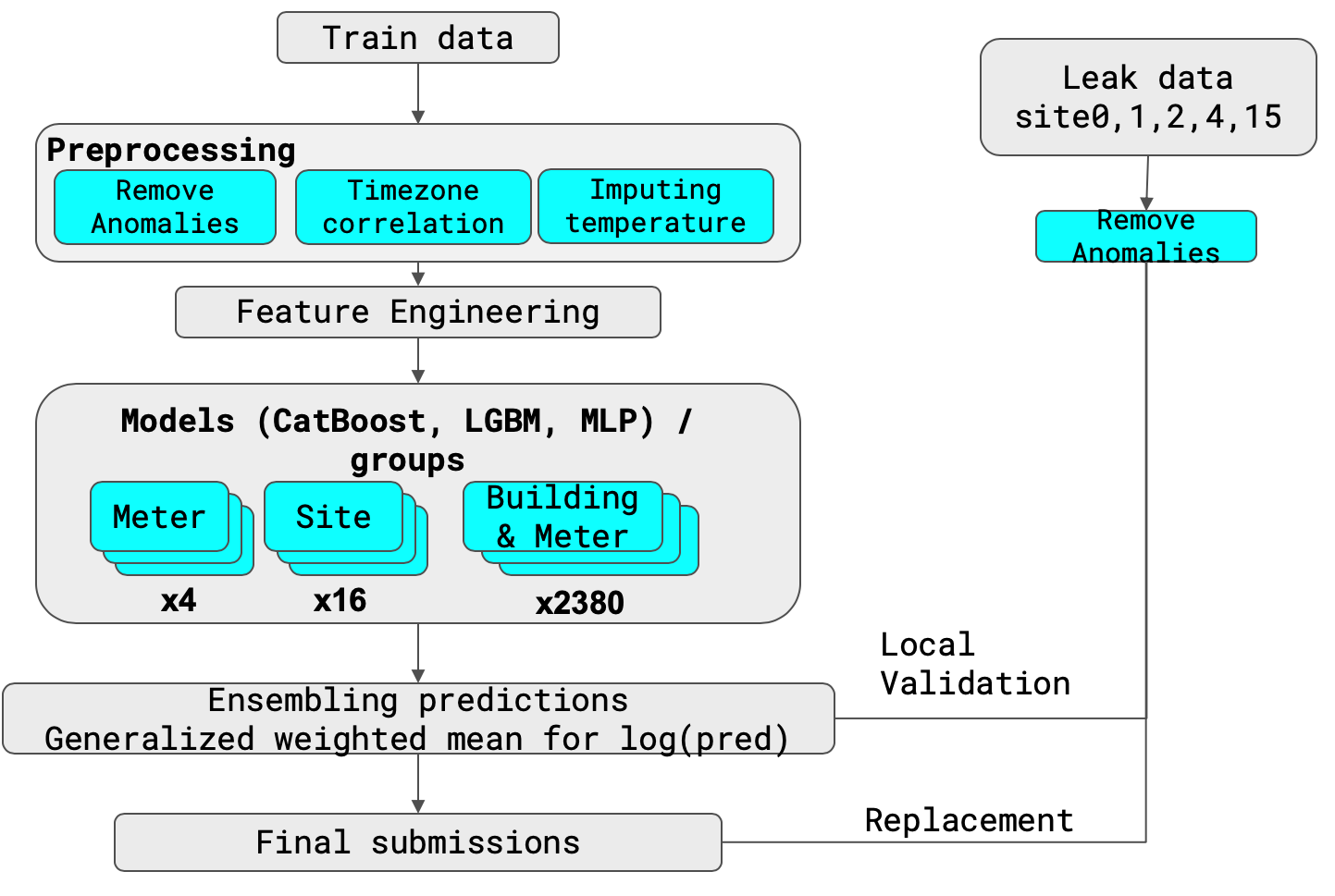}
\caption{Overview of the First-Ranked winning solution (Adapted from the winners documentation submission and used with permission.)}
\label{fig7-first}
\end{figure}

The workflow of the first-place team’s solution is shown in Figure \ref{fig7-first}. The first step of their solution involved preprocessing the given competition dataset by removing anomalies, imputing missing weather data values, and correcting time zones. Their next step involved feature engineering in which they extracted 28 features, such as raw data (e.g., meter, building metadata and weather parameters), categorical interactions between building metadata and meters, temporal (e.g., holidays, time of the day), various features of the weather data, and different target encoding features. Their model development strategy involved training separate models, one each on different subsets of the training dataset using three algorithms -- CatBoost \citep{Prokhorenkova2018-qk}, LightGBM \citep{Ke2017-ya}, and Muli-Layer Perceptrons (MLP). The training subsets were based on \texttt{meter}, \texttt{primary\_use}, and \texttt{site\_id}. All the models were tuned using a 12-fold cross-validation method by using the following consecutive eleven months data as the public test/validation set. Finally, individual model predictions were combined using a generalized weighted mean approach to obtain the final predicted values.

\subsection{Second-place solution - Intensive pre-processing and a huge XGBoost, LightGBM, CatBoost, and FFNN ensemble}

The second-placed team was made up of Rohan Rao, a Senior Data Scientist from Bangalore, India; Anton Isakin, a Machine Learning Engineer from Nurenberg, Germany; Yangguang Zang a Data Scientist from Beijing, China; and Oleg Knaub, a Data Scientist from Amberg, Germany. Their \emph{final private test score} (private leaderboard) was 1.232 and their \emph{public test/validation set score} (public leaderboard) was 0.937 with a rank of 12. Rohan was at the \emph{Grandmaster} level on the platform, and this was the third competition in which he had placed in the top 10. The other members of the team were either at the \emph{Master} or \emph{Expert} levels.

\begin{figure*}[ht]
\centering
\includegraphics[width=\textwidth,height=\textheight,keepaspectratio]{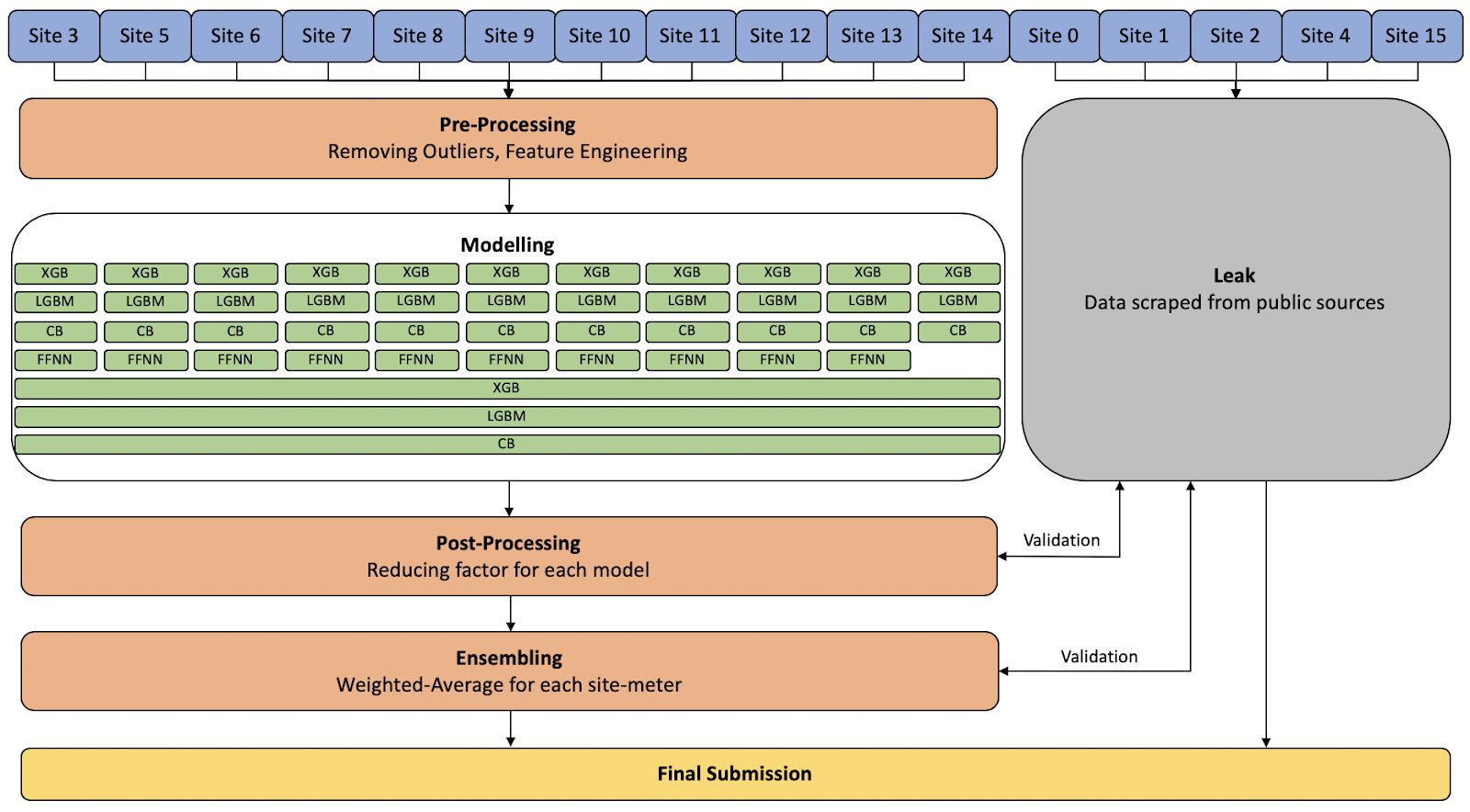}
\caption{Overview of the Second-Ranked solution (Adapted from the winner's documentation submission and used with permission.)}
\label{fig8-second}
\end{figure*}

The second-place solution overview is shown in Figure \ref{fig8-second}. This team removed the outlier data manually by visually inspecting and filtering each building data. Their feature selection involved calculating simple statistics of the weather and building metadata in addition to the temporal features, but there was no focus on using sophisticated lag features. Similar to the first place solution, this team also developed different models that were trained on different data subsets and the entire training set using XGBoost \citep{Chen2016-vx}, LightGBM \citep{Ke2017-ya}, Catboost \citep{Prokhorenkova2018-qk}, and Feed-forward Neural Networks (FFNN). FFNN models were built only for the meter id \texttt{0} (electrical meter). Finally, the predictions from individual models were ensembled using the weighted mean method, similar to the first-place team, and the weights were determined empirically.

\subsection{Third-place solution - Catboost and LightGBM with weighted post-processing}

The third-place winner of the competition was a single contestant named Xavier Capdepon, a Senior Data Scientist in New York City, NY. His \emph{final private test score} (private leaderboard) was 1.234 and his \emph{public test/validation set score} (public leaderboard) was 0.946 with a rank of 44. Xavier was at the \emph{Master} level on the platform and mentioned in discussion with the technical team that he has a background in Civil Engineering. 

\begin{figure}[ht]
\centering
\includegraphics[width=\linewidth]{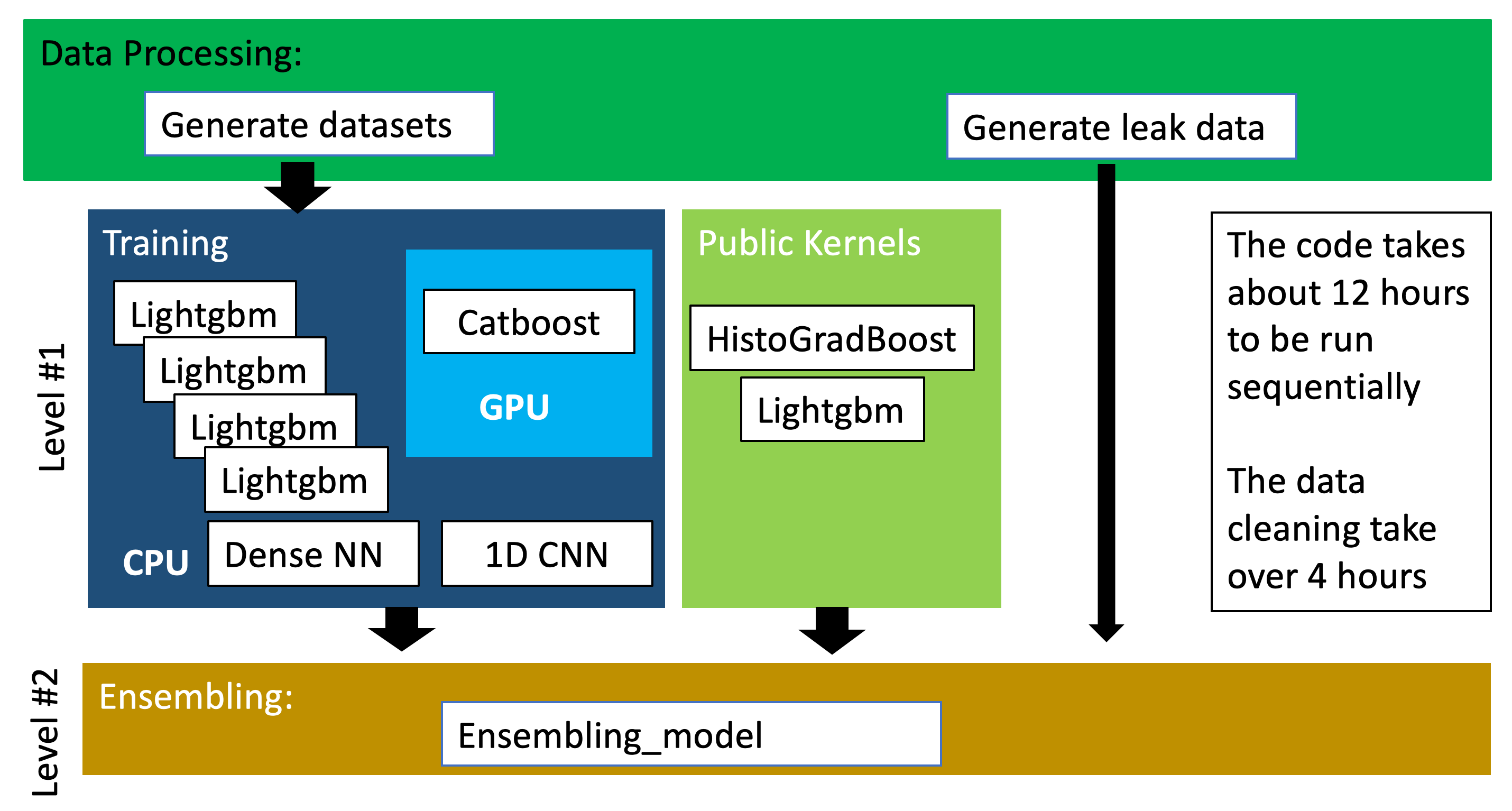}
\caption{Overview of the Third-Ranked Solution  (Adapted from the winner's documentation submission and used with permission.)}
\label{fig9-third}
\end{figure}

Similar to the first and second-place solution, Xavier also applied a log transformation to the target variable. His feature engineering strategy involved computing derived weather features, such as heat, windchill, and lagged weather features, in addition to the temporal and building metadata, resulting in 23 features in total. The model development process consisted of training a set of models that included Catboost \citep{Prokhorenkova2018-qk}, neural networks, and LightGBM \citep{Ke2017-ya} models. Finally, he ensembled the individual model predictions using a weighted average method where weights were calculated with the help of the publicly-available datasets. Figure \ref{fig9-third} illustrates an overview of the analysis process of Xavier’s solution.

\subsection{Fourth-place solution - XGBoost and LightGBM}
The fourth-place winner of the competition was a single contestant named Jun Yang, a student at the University of Electronic Science and Technology of China in Chengdu, China. Jun was an \emph{Expert} on the Kaggle platform. His \emph{final private test score} (private leaderboard) was 1.235 and his \emph{public test/validation set score} (public leaderboard) was 0.936 with a rank of 48.

\begin{figure*}[ht]
\centering
\includegraphics[width=\textwidth,height=\textheight,keepaspectratio]{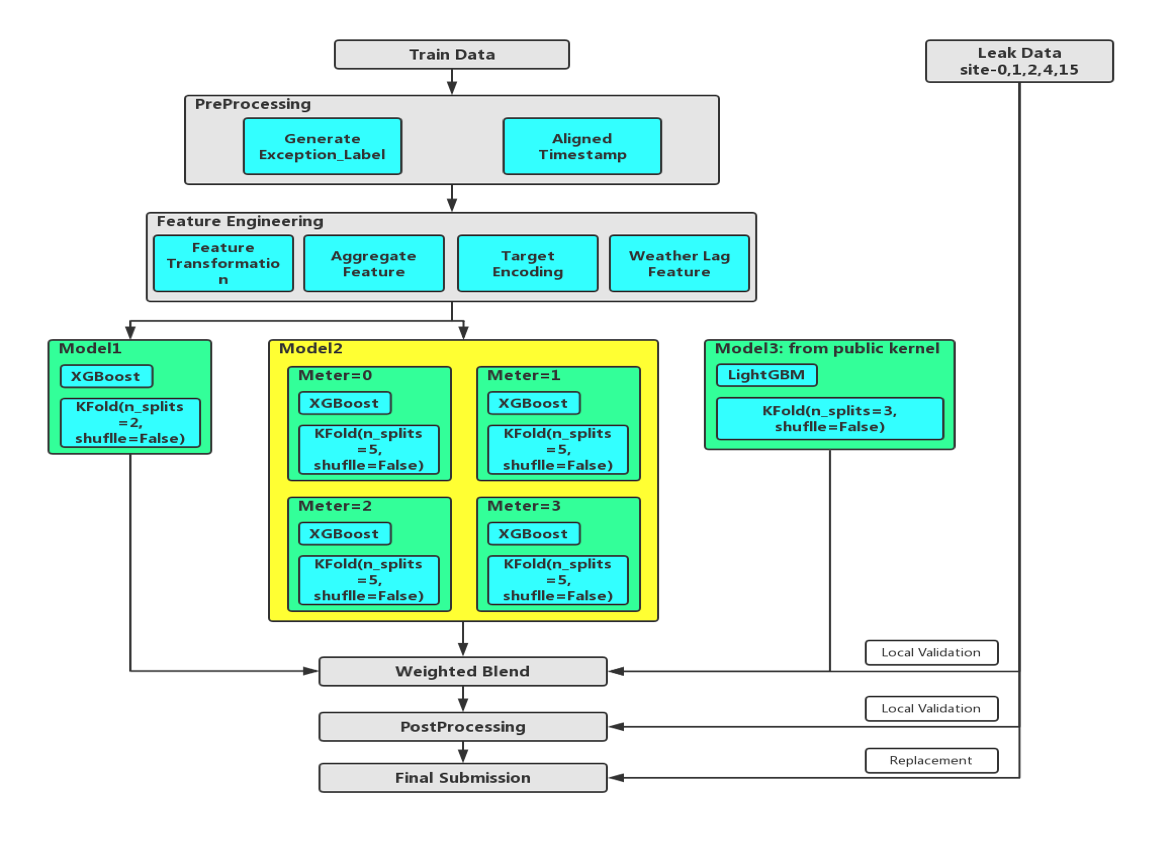}
\caption{Overview of the Fourth-Ranked solution (Adapted from the winner's documentation submission and used with permission)}
\label{fig10-fourth}
\end{figure*}

Like the first-place teams, Jun also extracted detailed features such as raw meter data, weather data, building metadata, temporal features from meter data, and lag features from raw weather data. An overview of his solution can be seen in Figure \ref{fig10-fourth}. The target variable was log-transformed before training the models. Two XGBoost models (using 2-fold and 5-fold cross-validation) \citep{Chen2016-vx} were trained using the whole training set. In addition to this, one LightGBM \citep{Ke2017-ya} model with 3-fold cross-validations, as extracted from Kaggle’s public kernels, was also trained using the whole training set. However, no systematic approach was followed to select the optimal hyperparameters. Some data cleaning was performed during each fold. The final prediction was made by ensembling all three models where weights were determined using the leaked data.

\subsection{Fifth-place solution - LightGBM ensemble with post-processing weighting}

The fifth-place winning team of the competition was made up of Tatsuya Sano, Minoru Tomioka, and Yuta Kobayashi, who are all students at the University of Tsukuba in Tsukuba, Japan. Their \emph{final private test score} (private leaderboard) was 1.237 and their \emph{public test/validation set score} (public leaderboard) was 0.940 with a rank of 23. 

\begin{figure}[ht]
\centering
\includegraphics[width=\linewidth]{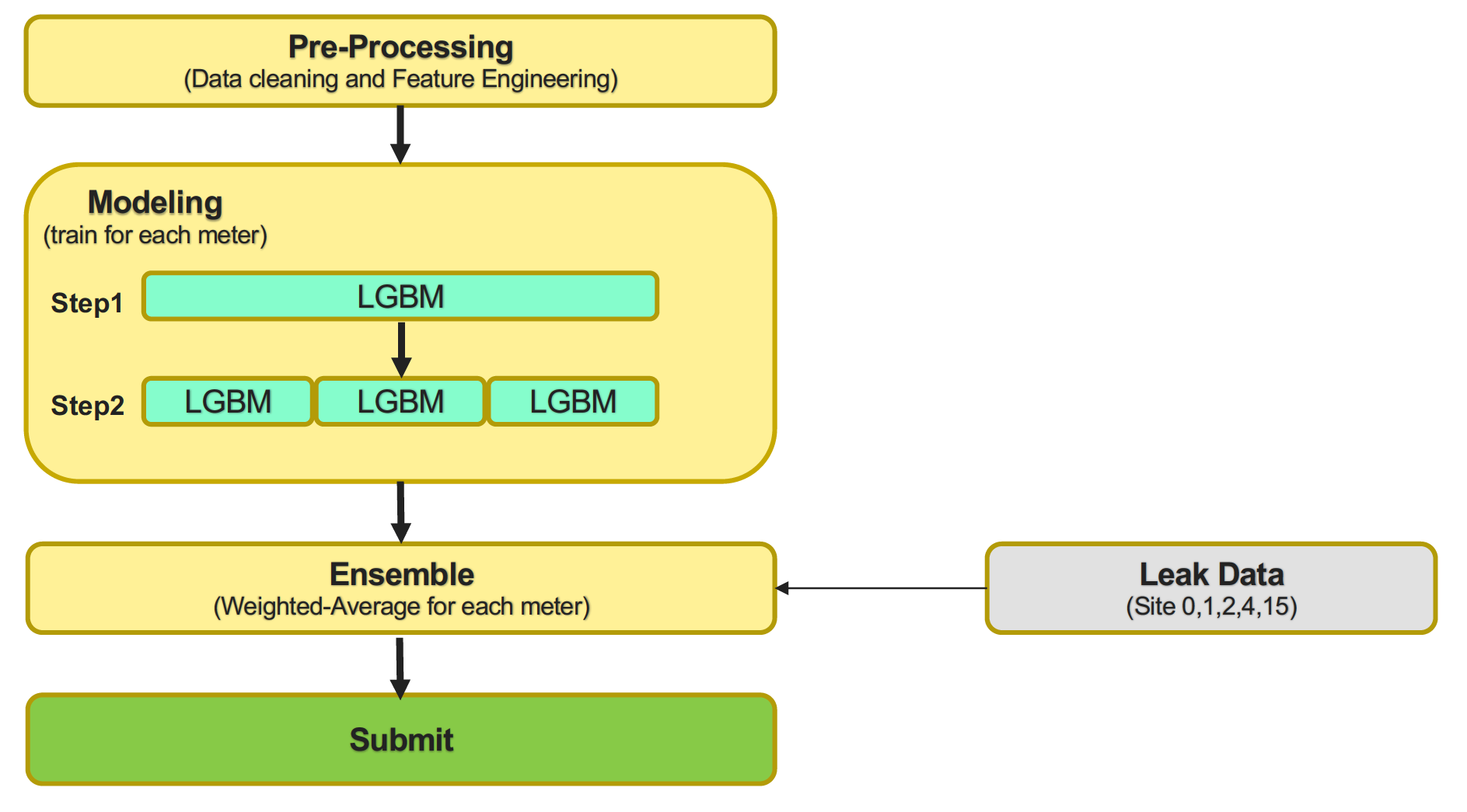}
\caption{Overview of the Fifth-Ranked solution (Figure adapted from the winner's documentation submission and used with permission)}
\label{fig11-fifth}
\end{figure}

The workflow of their solution is shown in Figure \ref{fig11-fifth}. In addition to focusing on different data preprocessing approaches, this team experimented with varying transformations of the target (percentile and proportion based). Unlike the other solutions, this team used only LightGBM \citep{Ke2017-ya} applied in a two-step process. In the first step, separate models were developed for each building and meter to determine one of the model parameters, number of trees, using a subset of the training set. In the next step, this optimal parameter was used to train separate models for each building and meter. Finally, individual model predictions were ensembled using a weighted average where the weights were determined based on the public dataset.

\subsection{Other medal winners}
\begin{figure}[ht]
\centering
\includegraphics[width=\linewidth]{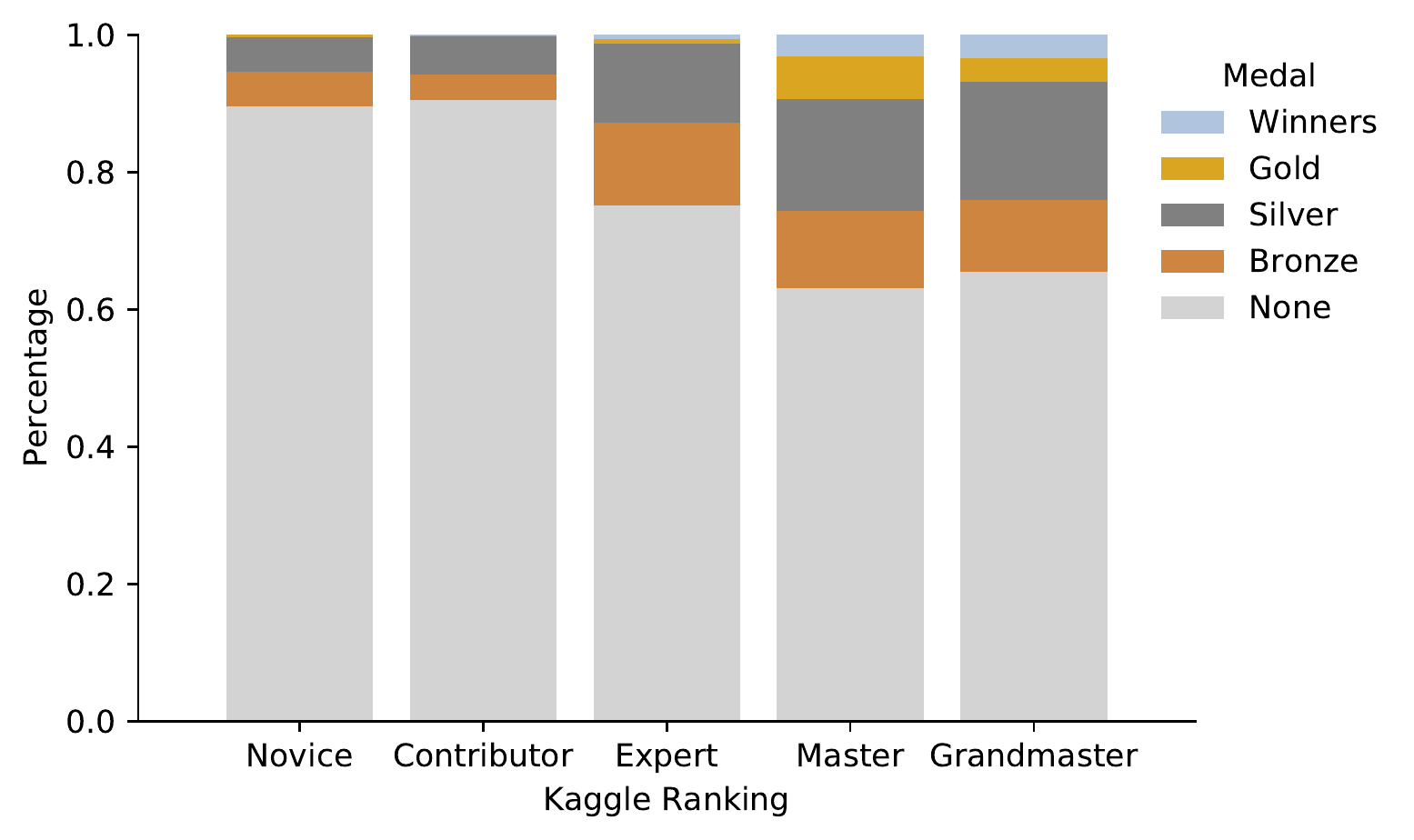}
\caption{Breakdown of the medal winners from the competition.}
\label{fig12-breakdown}
\end{figure}

In addition to the top five winners, other ranges of contestants were able to win medals based on their final standing in the competition. Figure \ref{fig12-breakdown} illustrates an overview of the medals that were won by the participants in this competition. The contest platform awards Gold (top 0.2\%), Silver (top 5\%), and Bronze (top 10\%) medals to contestants based on where their \emph{final private test score} (private leaderboard) rankings placed them at the end of the competition. Although the \emph{Novice} and \emph{Contributor} groups had the most participants, they observed a high percentage of participants who did not earn any medals (about 90\%). Participants with a \emph{Grandmaster} ranking were numbered at 29 - a much smaller demographic than other rankings - but about a third of these participants were awarded a medal of some kind. This situation was unsurprising since \emph{Grandmasters} have much more experience in machine learning and using the platform. 

\section{Shared machine learning workflows using notebooks}
Beyond the top winning teams, there were numerous participants throughout the competition timeline that created analysis examples to share with the public. While this may seem counter-productive towards winning the competition, a strong sense of community spirit and sharing occurred, especially in the early phases. The competitors saw sharing as a means of enhancing the final result by providing a level playing field for some of the more mundane tasks in the machine learning process, such as detecting erroneous data. These shared workflows were in a format known as a \emph{Notebook}, which gives participants the ability to share analysis code, instructions, explanation, and other forms of content in a single, shareable page.

\subsection{Overview of the notebook topics}
\begin{figure*}[ht]
\centering
\includegraphics[width=\textwidth,height=\textheight,keepaspectratio]{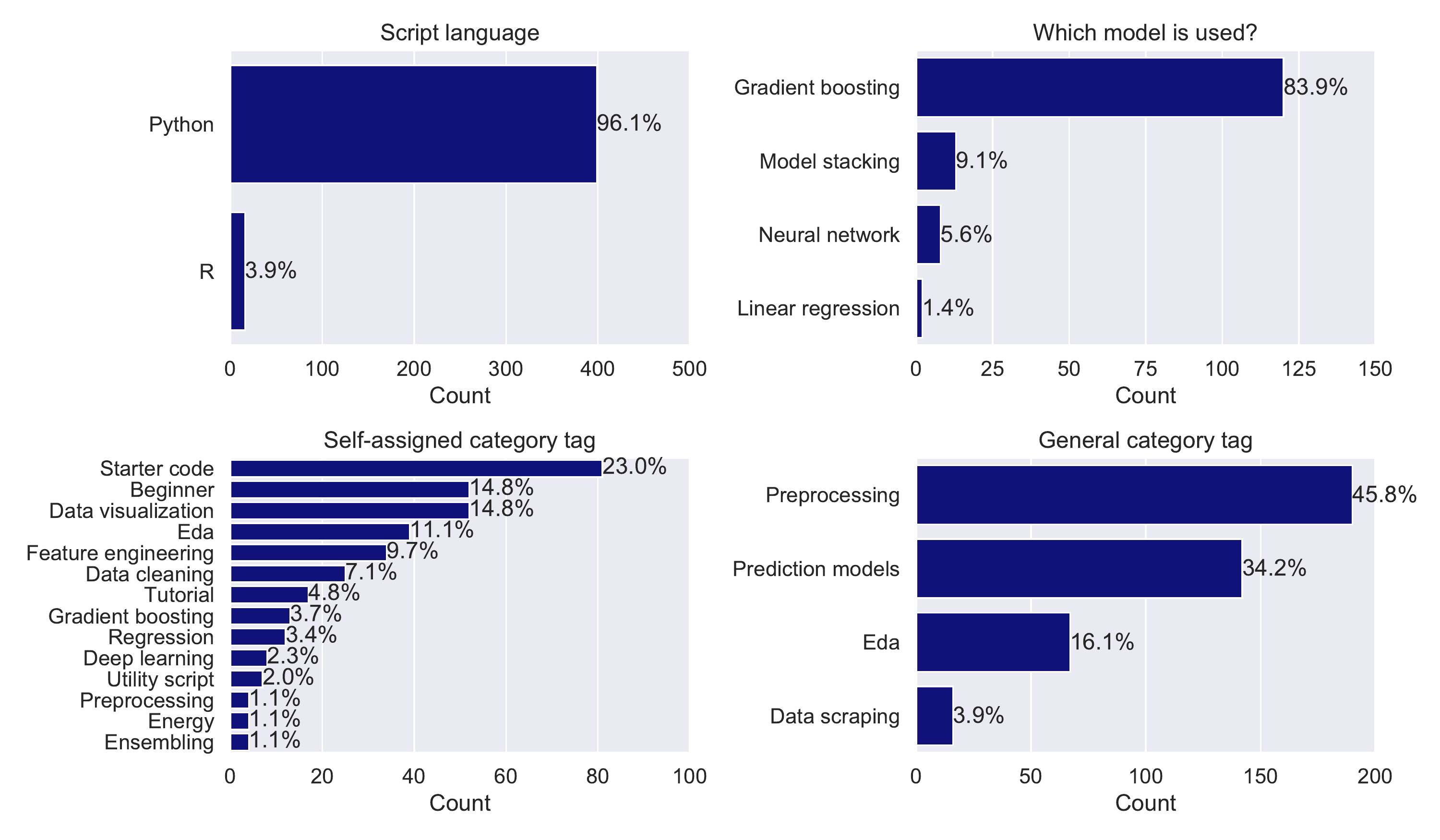}
\caption{Overview of Analysis Notebooks in the Competition: programming language (top left), model category (top right), general category tag (bottom right), and self-assigned category tag (bottom left)}
\label{fig13-notebooks}
\end{figure*}

Figure \ref{fig13-notebooks} provides a meta-analysis of 415 notebooks that were shared by competition participants. The goal of this analysis was to guide building science analysts on where to start if they would like to learn from the content created by the competition. The first category was the programming language used in the various notebooks. Python\footnote{\url{https://www.python.org/}} and R\footnote{\url{https://www.r-project.org/}} were the only two programming languages used, where Python was chosen by over 96\% of the competitors. The following breakdown focused on the use of specific machine learning models. Gradient boosting trees dominated with almost 84\% of the notebooks using this model type. Neural networks and linear regression were two other specific model types that were used, and the category model stacking signifies notebooks that used ensembles of models.

We also examined the general categories of the analysis notebooks by manually assigning tags to each one. The largest group was related to the preprocessing of data, with almost half of the notebooks dedicated to this topic. Preprocessing emerged as a critical aspect of the performance of the winning models. Prediction models themselves accounted for a third of the notebooks, with the rest split between exploratory data analysis (EDA) (16.1\%) and data scraping (3.86\%). Figure \ref{fig13-notebooks} analyzes the self-assigned tags that the users gave their notebooks upon posting them. These labels include descriptors for aspects, such as whether the notebooks are suitable for beginners or as starter code. This tag descriptor breakdown includes technical descriptors such as deep learning, ensembling, and feature engineering.

\subsection{Complete reproducible machine learning solution examples}
In addition to the top five winning teams, we were able to extract full machine learning workflows from the notebooks and discussion boards that were openly shared by many of the competitors. Table \ref{table-solutions} shows a breakdown of these solutions. This table is meant to serve as a directory for future users who want to learn about specific types of machine learning and data analysis.

\begin{table*}[ht!]

\caption{List of Full Solutions found in the Notebooks.\\
Preprocessing key: RO = Remove Outliers; I = Imputation\\
Feature strategy key: H = Holiday Features; CS = Catgorical Statistical features\\
Modeling strategy key: LGBM = LightGBM; CB = CatBoost; NN = Neural Network; RF = Random Forrest; LR = Linear Regression; XG = XGBoost; LM = LightMORT; The parenthesis after each model indicate how many models are used in the ensembles.\\
Post-processing key: EM = Ensemble Model}
\label{table-solutions}
\centering

\begin{tabular}{|p{0.7cm}|p{3.5cm}|p{0.7cm}|p{1.4cm}|p{1.4cm}|p{1.1cm}|p{3.9cm}|p{1cm}|}
\hline
\textbf{Rank} & \textbf{Team Name} & \textbf{Score} & \textbf{Pre-process} & \textbf{Feature Strategy} & \textbf{Features} & \textbf{Modeling Strategy} & \textbf{Post-Process} \\
\hline
9 & MPWARE & 1.241 & RO, I & H, CS & - & LGBM(7), CB(4), LM(1), NN(4) & EM \\
\hline
13 & Tim Yee & 1.243 & RO, I & CS & - & LGBM(3) & EM \\
\hline
20 & [ods.ai] PowerRangers & 1.244 & RO, I & - & - & LGBM(15), NN(5) & EM \\
\hline
25 & Georgi Pamukov & 1.245 & RO, I & H & - & LGBM, NN, LR & EM \\
\hline
46 & Fernando Wittmann & 1.248 & - & - & - & LGBM(9) & EM \\
\hline
52 & Pavel Gusev & 1.25 & - & - & - & LGBM(4) & EM \\
\hline
77 & CR7 & 1.256 & RO, I & H, CS & 50 & LGBM(10), CB(1) & EM \\
\hline
173 & Electrium Z & 1.267 & RO, I & H, CS & - & LGBM, CB, XG, RF, NN & EM \\
\hline
367 & patrick0302 & 1.28 & RO, I & H, CS & 40 & LGBM(7) & EM \\
\hline
497 & Hiroyuki Namba & 1.286 & RO & - & 18 & LGBM(32) & EM \\
\hline
1545 & KottayamKings & 1.31 & RO & H & 14 & LGBM (1) & - \\
\hline
1678 & Taegwan Kim & 1.32 & - & H & 14 & LGBM (1) & - \\
\hline
1703 & Vishwanath R Kulkarni & 1.326 & RO, I & CS & 25 & LGBM (1) & - \\
\hline
1710 & Georgios Chatzis & 1.327 & RO & - & 17 & LGBM (1) & - \\
\hline
1727 & UniTartu\_ML & 1.332 & RO, I & CS & 27 & LGBM (1) & - \\
\hline
1866 & Sergei Tsimbalist & 1.369 & RO, I & - & 18 & LGBM (1) & - \\
\hline
1920 & Hitesh Somani & 1.378 & RO, I & - & 24 & LGBM (1) & - \\
\hline
2058 & Clement\_ut & 1.385 & RO, I & CS & 27 & LGBM (1) & - \\
\hline
2066 & Atharva Patel & 1.386 & RO & H & 24 & LGBM (1) & - \\
\hline
2082 & CodeNinja & 1.387 & RO, I & - & 22 & LGBM (1) & - \\
\hline
2130 & Ergo Nigola & 1.39 & RO, I & CS & 29 & LGBM (1) & - \\
\hline
2163 & mevrick & 1.393 & RO, I & H & 21 & LGBM (1) & - \\
\hline
2232 & Viswajith & 1.397 & RO, I & H & 16 & LGBM (1) & - \\
\hline
2373 & MANISH SHUKLA & 1.41 & RO, I & - & 21 & LGBM (1) & - \\
\hline
2400 & Stéphane Thibaud & 1.412 & RO, I & - & 15 & LGBM (1) & - \\
\hline
2538 & luisfer & 1.436 & I & - & 13 & NN(1) & - \\
\hline
2549 & Srinivas M Besthar & 1.439 & I & - & 22 & LGBM (1) & - \\
\hline
2589 & Vikas Singh & 1.452 & I & - & 21 & LGBM (1) & - \\
\hline
2646 & Ishaan Jain & 1.47 & I & - & & LGBM (1) & - \\
\hline
2761 & Gouher Danish & 1.526 & RO, I & - & 17 & LGBM (1) & - \\
\hline
2767 & Paul Larmuseau & 1.529 & I & - & 23 & LR & - \\
\hline
2801 & Aldrin & 1.549 & I & - & 20 & RF(1) & - \\
\hline
3077 & Pierre-Matthieu Pair & 1.839 & I & - & 13 & LGBM (1) & - \\
\hline
3258 & Sreelatha Renukuntla & 2.265 & - & - & 19 & RF(1) & - \\
\hline
3296 & Sneaky Weasels & 2.393 & RO, I & - & 4 & RF(1) & - \\
\hline
3372 & Ceren Iyim & 2.704 & I & CS & 24 & LGBM (1) & - \\
\hline
3514 & Jakub Ciborowski & 4.21 & - & - & 21 & LGBM (1) & - \\
\hline
\end{tabular}

\end{table*}

\section{Discussion}
This competition uncovered several key insights into the scalability of machine learning for the energy prediction context. This section gives an overview of the general knowledge gained from the competition that can be utilized and expanded upon by the data-driven building energy prediction domain.

\subsection{Objective judgment of model types and steps for building energy prediction}

The competition's primary goal was to be able to compare numerous configurations of modeling techniques developed by thousands of machine learning practitioners, hundreds of them being experts in the field. Several vital observations became apparent through the analysis of the top five winning solutions as well as the dozens of whole machine learning solution examples. These insights can be considered generalizable for the type of data provided in the competition. 

\subsubsection{Pre and postprocessing steps are essential and can’t always be automated}

One of the key differences between the top-performing solutions and solutions further down the leaderboard was the methods that the contestants used to filter the data before modeling and the corrections they applied after prediction and before submission. For example, the top two winning teams explained that they spent a significant amount of time removing anomalous behavior from the dataset. The second-place solution described how they ended up manually doing this process through visualization for all meters; this insight was informative for a team with \emph{Grandmaster} and \emph{Master} machine learning experts. This insight indicates the importance of domain understanding and the difficulty in fully automating the process of removing data that adds false signals to the training data. For postprocessing, the top winners also included various methods to apply weightings and created complex ensembling frameworks to develop high accuracy with such a high amount of training data.

\subsubsection{Gradient boosting tree models dominate}
One of the most prominent outcomes of the competition was the domination of tree-based machine learning models, specifically gradient boosting trees. This type of model is prevalent in many machine learning competitions, and it proved to be the most commonly applied to this context. It was a part of all winning solutions, and most of the posted analysis notebooks from the rest of the leaderboard. These models are popular because they provide high accuracy and flexibility that work well with numerous types of data. The most commonly used model was the LightGBM framework that is designed to be fast in training speed and low in memory usage, while still maintaining high accuracy\footnote{\url{https://github.com/microsoft/LightGBM}}. Other gradient boosting models used include CatBoost\footnote{\url{https://catboost.ai}}, XGBoost\footnote{\url{https://github.com/dmlc/xgboost}}, and LiteMORT\footnote{\url{https://github.com/closest-git/LiteMORT}}. A few other models were used in the large winning ensembles such as Multi-Layer Perceptrons, Feed-forward Neural Networks (FFNN), and Random Forest models.

\subsubsection{Ensembles of models create more accurate (but slower) solutions}
An analysis of the top five solutions showed that more complex ensemble-based models were necessary to create the most accurate solutions. The large ensemble methods used created numerous models that were trained on various subsets of the dataset, such as the type of building or meter. Structuring models in this way creates stronger results through diversity of models that reduce bias and variance by incorporating different estimators with different patterns of error. Despite their superiority in terms of accuracy, large ensembles of complex models like gradient boosting trees can be computationally intensive and can take hours, or even days to complete the training and prediction process. The trend of using ensemble methods was also identified in the publications of building science research field \citep{Wang2017-yn}. A worthy area of further analysis for this competition is to compare the complexity versus the practical use trade-off for the winning models versus those found further down the leaderboard. 

\subsection{Convergence of skills and talent from the engineering and data science domains}
It was apparent that a majority of the participants were not members of the building science or research community. Most contestants who were in the top medal-winning levels of the leaderboard tended to be at least at the \emph{Expert} machine learning level, however, there were several exceptions. Despite the large number of data science experts, there were some examples of engineering knowledge that aided the  winning solutions. A member of the top team and the third place winner had an engineering background that was relevant to the competition context.

\subsection{Dataset creation and preparation}
The amount of networking needed to collect enough datasets for the competition was higher than expected. Despite a wide-spread request for data at the ASHRAE Winter Meeting in Atlanta in January 2019, the majority of the datasets either came from personal contacts of the technical team or from publicly available online sources. Another major challenge for the technical team was creating a consistently formatted and organized schema derived from all the various data sources. Several issues related to mistakes in cleaning became apparent after the start of the competition. For example, one of the datasets had inconsistent units and wasn't properly converted to kWh. Another situation was that the weather data had a time-stamp that was not aligned with the local time for the individual meters. In both of these cases, the winning competitors were able to recognize the issue and crowdsource a solution. The community was able to create publicly available notebooks that helped level the competitive playing field despite the issues.

\subsection{Inclusion of publicly available data in the competition}
The technical team's primary goal in preparing for the competition was the development of a big enough data set to make the competition viable. Having a large and diverse dataset ensures that the winning solutions will generalize well to many other types of buildings located around the world. The decision was made early in the process to include the use of publicly available data due to its availability and the value that it can provide for contestants in building models. The planning team's strategy was to protect the integrity of evaluating the winning team’s solution using only non-public data for use in the \emph{final private test data set} (private leaderboard), which would be the determining factor for the winning teams. The publicly available data would only be used for the \emph{public test/validation data set} (public leaderboard) portion of the competition. This situation would allow for the use of a larger dataset while protecting the integrity of evaluating the winning models. It is not uncommon in machine learning competitions to have different types of hold out data sets, with one focused on being used as a tool to gauge short-term success and another to serve as the final deciding set to determine winners.

The competitors were not initially notified that portions of the \emph{public test/validation data set} (public leaderboard) were openly available online, but the competitors quickly found many of the datasets and introduced them as \emph{leakage data} in the discussion boards. Although it was expected that these public data sets would be discovered, the negative tone on discussion topics about this issue was not expected; the competitors felt that using these data was unfair and that the public leaderboard became less useful. This frustration was diminished for most of the participants, however, as open-sourced scraping methods were shared on discussion threads to collect and disseminate these data. All publicly-available data sets discovered by the contestants were shared openly and no contestant had an \emph{unfair} advantage due to these data. In fact, there were many creative uses of the publicly-available data that enriched the competition by providing more potential for \emph{validation data} set generation. Not all of the public data sets were discovered, and the winning solutions did not use all the publicly available data that was uncovered by participants on the discussion boards.

While the contestants considered the publicly available data sets in the \emph{public test/validation data set} (public leaderboard) as leakage, these data were not used to determine the winner. The winning teams were judged based on their performance on the \emph{final private test data} (private leaderboard). The only leakage that occurred concerning this data set was a single site that was discovered to have data available online for which the competition planning committee was unaware. This data set was switched to the public leaderboard data set towards the end of the competition, and its impact was removed from the resulting competition scoring at the conclusion.

Future machine learning competitions should avoid using publicly available data in even the public leaderboard component of the structure based on the experience of the GEPIII competition planners. This element produced an amount of frustration that was more disruptive than what benefit the data provided. If public data is included, the contestants should be notified at the beginning of the competition of their presence.

\subsection{Generalizability of the competition results}
While this competition was a significant step in the direction of the comparability of machine learning techniques for building energy prediction, there are still several limitations to its results in application to practice and research. The primary constraint is the diversity of the data set created for the competition; most of the buildings are from the education sector, namely higher education such as university campuses. The resultant models developed are therefore fitted on this type of data, and using these models on other types of building data from different sectors remains untested at this time. The goal of this competition was to compare techniques objectively and use extensive, open data to do that. A lot of potential remains for other entities to grow the data set used in this competition and test whether the developed techniques need various tuning or modifications to be more generalizable. This effort will likely need to be community-driven, with data disclosure for buildings becoming the norm in cases where personal privacy is not an issue. In addition, there are more challenging machine learning applications such as prediction of peak loads or time and seasonal differentiation of savings that could be the objective and be tested in the context of a competition. The planning team foresees a future Great Energy Predictor competition could include an even larger data set with a better representation of the different building types and climates worldwide and applied to more complex objectives that could better serve the application needs of the domain.

\section{Conclusion}
The competition's objective was to create value for building industry practitioners who want to improve the prediction models used for building performance analysis. This competition was successful at creating the most extensive crowdsourced machine learning solution and benchmarking exercise in the building energy research domain. Users of these techniques can be confident that the accuracy of the models has been verified as compared to tens of thousands of alternative models developed by numerous experts. There are several areas of targeted output for the broader community to learn from, especially those that are just starting in data science to disseminate these models. 

\subsection{A reproducible repository of the winning solutions}
The code and instructions for execution for the top five winning solutions have been posted in a Github along with guidelines of how to reproduce and use them in various applications in the built environment\footnote{\url{https://github.com/buds-lab/ashrae-great-energy-predictor-3-solution-analysis}}. These solutions also include less complicated versions that allow for users to decide between the best complexity to accuracy balance they are trying to achieve. This repository consists of the original code that was submitted by the winning contestants, their submitted solution summary document, a link to the explanation video provided by the winning team, and detailed instructions on how to reproduce each solution including information about resources, run time, and computational effort\footnote{\url{https://github.com/buds-lab/ashrae-great-energy-predictor-3-solution-analysis/wiki}}.

\subsection{Open-source building energy benchmarking data}
The training data set (2016) for this competition is still available for download and use from the Kaggle competition website\footnote{\url{https://www.kaggle.com/c/ashrae-energy-prediction}}. Researchers can even upload predictions as a \emph{Late Submission} and see where their predictions fall as compared to the public test/validation data set (public leaderboard) and final private test data (private leaderboard). The technical team for the competition has released a more extensive data set that includes the training (2016) and public test/validation (2017) data from this competition in addition to numerous other buildings\footnote{\url{https://github.com/buds-lab/building-data-genome-project-2}}. This data set is released as the Building Data Genome Project 2 (BDG2) \citep{Miller2020-hc}, the next iteration of the BDG \citep{Miller2017-fq}. This data set is also available as a sandbox in the Kaggle data page\footnote{\url{https://www.kaggle.com/claytonmiller/building-data-genome-project-2}}.

\subsection{Curated directory of machine learning workflow examples}
This paper has outlined the key categories of data science knowledge created through publicly-shared Kaggle notebooks. A continuously updating wiki has been set up to create a venue for notebooks from this competition to be tagged and posted according to various categories\footnote{\url{https://github.com/buds-lab/ashrae-great-energy-predictor-3-overview-analysis/wiki/Curation-of-Machine-Learning-Tutorials}}. This wiki outlines the best notebooks for tutorials in which building science professionals can use to expand their skill set. This wiki is part of a repository that includes the data and code used to create Figures \ref{fig1-data}-\ref{fig6-discuss} and \ref{fig12-breakdown}-\ref{fig13-notebooks} as well as other notebooks describing how the data was collected for this publication\footnote{\url{https://github.com/buds-lab/ashrae-great-energy-predictor-3-overview-analysis}}.

\subsection{Future workshops and seminars}
Several ASHRAE Seminars and workshops were planned to disseminate the results of the GEPIII competition. These seminars will allow the winners to meet and address the ASHRAE community, to transfer knowledge, and to learn more about the building science community. A conference track devoted to an overview of the top three winning solutions was held at the ASHRAE Summer Meeting held online in June 2020.

\subsection{Future machine learning competitions}
\subsubsection{CityLearn challenge}
Beyond data-driven energy prediction methodologies, as developed in this competition, are methods and algorithms for direct energy management and control. The major challenge is the scalability of these approaches across many buildings, given their unique characteristics. Reinforcement learning (RL) has gained popularity in the research community as a model-free and adaptive controller for the built environment. RL has the potential to be an inexpensive plug-and-play controller that can be easily implemented in any building, regardless of its model (unlike model predictive controllers or MPCs), and coordinate multiple buildings for applications, such as demand response and load shaping. Related to these motivations, the CityLearn Challenge was held in early 2020 \citep{Vazquez-Canteli2019-rr}. CityLearn is an OpenAI Gym environment for the implementation of RL agents for demand response at the urban level. The environment allows the implementation of single-agent (as a centralized agent) and multi-agent decentralized RL controllers. 

\subsubsection{Great Predictor IV competition}
Plans are being made to continue the ASHRAE Great Predictor Machine Learning competitions with the possible objective of time-series classification focused on building performance or operations data from building management and automation systems. This objective will likely include the use of millions of data streams of building management system data from thousands of buildings. The focus of this prediction will focus on the automation of meta-data inference of these systems such that the implementation of energy savings calculation techniques become more scalable, and new markets are created due to this lower barrier.

\section*{Acknowledgements}
The authors acknowledge the individuals who assisted in the collection of the data set for this competition. This list includes (alphabetical order) Adam Boltz, Adam Keeling, Ann Lundholm, Araz Ashouri, Brodie W. Hobson, Catherine Patton, Doug Livingston, Gerry Hamilton, Ian Lahiff, James Ball, Jonathan Roth, Justin Owen, Kian Wee Chen, Maxime St-Jacques, Nate Boyd, Paul Raftery, Saptak Dutta, Zach Wilson, and Zixiao Shi. The Kaggle platform technical and advisory team, including Addison Howard and Sohier Dane, were instrumental to the launch of the Great Energy Predictor III competition. The authors would also like to the thank the individuals and teams from the top five winning submissions for their time and support in explaining their solutions. This group includes Matthew Motoki, Isamu Yamashita, Rohan Rao, Anton Isakin, Yangguang Zang, Oleg Knaub, Xavier Capdepon, Jun Yang, Tatsuya Sano, Minoru Tomioka, and Yuta Kobayashi.

Financial support for the competition monetary prizes was supported by ASHRAE. The Kaggle machine learning platform provided hosting as a non-profit competition. Additional financial support for the development of the competition and travel support was provided by the Republic of Singapore's National Research Foundation through a grant to the Berkeley Education Alliance for Research in Singapore (BEARS) for the Singapore Berkeley Building Efficiency and Sustainability in the Tropics (SinBerBEST) program. Additional support was provided by the Ministry of Education (MOE) of the Republic of Singapore (R296000181133).

\section*{Author contributions}
C.M. was the chair of the technical committee, led the data collection, analysis and deployment of the competition, and was the lead and corresponding author of the paper. P.A. was a member of the technical committee, co-led the data preparation and model prototyping process, and validated the winning solutions. A.K. was a member of the technical committee and co-led the data preparation process and post-competition analysis. J.Y.P. and Z.N. were members of the technical committee and assisted in the data preparation, launch and some post-competition analysis. C.F. and J.R. assisted in the post-competition and results analysis. C.B. was the chair of the competition planning committee and assisted in the preparation of the paper. K.G., A.F., and J.H. were members of the competition planning committee and assisted in the preparation, launch, and the discussion of the results. All authors reviewed the manuscript and take responsibility for its content.

\section*{Competing interests}

The authors declare no competing interests.

\bibliographystyle{model1-num-names}
\bibliography{00_main}

\end{document}